\documentclass[prb,twocolumn,amsmath,amssymb]{revtex4}

\usepackage{graphicx}
\usepackage{amsmath}
\usepackage{amssymb}
\usepackage{amsfonts}
\usepackage{graphicx}
\usepackage{color}
\usepackage{mathrsfs}
\usepackage{epsf}
\usepackage{verbatim}
\usepackage{natbib}
\usepackage[hypertex]{hyperref}

\newcommand{\z}{\negthickspace}
\newcommand{\al}{\alpha}
\newcommand{\be}{\beta}
\newcommand{\ga}{\gamma}
\newcommand{\de}{\delta}
\newcommand{\ep}{\epsilon}

\newcommand{\pr}{^{\prime}}

\newcommand{\nn}{\nonumber}

\allowdisplaybreaks[1]

\newcommand{\ov}{\overline}

\newcommand{\1}{r_1}
\newcommand{\2}{r_2}
\newcommand{\3}{r_3}
\newcommand{\4}{r_4}
\newcommand{\q}{r_q}

\newcommand{\ra}{\rangle}
\newcommand{\la}{\langle}

\newcommand{\dg}{^{\dagger}}

\newcommand{\bb}{\bar{b}}
\newcommand{\fb}{\bar{f}}

\newcommand{\bph}{b^{\vphantom{\dagger}}}
\newcommand{\fph}{f^{\vphantom{\dagger}}}

\newcommand{\bbd}{\bb^{\dagger}}
\newcommand{\fbd}{\fb^{\dagger}}

\newcommand{\bd}{b^{\dagger}}
\newcommand{\fd}{f^{\dagger}}

\newcommand{\cd}{c^{\dagger}}
\newcommand{\dd}{d^{\dagger}}
\newcommand{\pd}{\partial}

\newcommand{\cph}{c^{\vphantom{\star}}}

\newcommand{\bbph}{{\bar b}^{\vphantom{\dagger}}}
\newcommand{\fbph}{{\bar f}^{\vphantom{\dagger}}}

\newcommand{\dph}{d^{\vphantom{\dagger}}}

\newcommand{\up}{\uparrow}
\newcommand{\down}{\downarrow}

\newcommand{\bup}{\bph_\up}
\newcommand{\fup}{\fph_\up}
\newcommand{\bdup}{\bd_\up}
\newcommand{\fdup}{\fd_\up}
\newcommand{\bdown}{\bph_\down}
\newcommand{\fdown}{\fph_\down}
\newcommand{\bddown}{\bd_\down}
\newcommand{\fddown}{\fd_\down}
\newcommand{\bbup}{\bbph_\up}
\newcommand{\fbup}{\fbph_\up}
\newcommand{\bbdup}{\bbd_\up}
\newcommand{\fbdup}{\fbd_\up}
\newcommand{\bbdown}{\bbph_\down}
\newcommand{\fbdown}{\fbph_\down}
\newcommand{\bbddown}{\bbd_\down}
\newcommand{\fbddown}{\fbd_\down}

\begin{document}

\bibliographystyle{revtex}

\title{
Boundary criticality and multifractality at the 2D spin quantum
Hall transition}

\author{Arvind R. Subramaniam}

\affiliation{James Franck Institute and Department of Physics,
University of Chicago, 5640 S. Ellis Ave., Chicago IL
60637, USA}

\author{Ilya A. Gruzberg}

\affiliation{James Franck Institute and Department of Physics,
University of Chicago, 5640 S. Ellis Ave., Chicago IL
60637, USA}

\author{Andreas W. W. Ludwig}

\affiliation{Department of Physics, University of California,
Santa Barbara CA 93106, USA}

\begin{abstract}

Multifractal scaling of critical wave functions at a
disorder-driven (Anderson) localization transition is modified
near boundaries of a sample. Here this effect is studied for
the example of the spin quantum Hall plateau transition using
the supersymmetry technique for disorder averaging. Upon
mapping of the spin quantum Hall transition to the classical
percolation problem with reflecting boundaries, a number of
multifractal exponents governing wave function scaling near a
boundary are obtained exactly. Moreover, additional exact
boundary scaling exponents of the localization problem are
extracted, and the problem is analyzed in other geometries.
\end{abstract}

\date{July 25, 2008}

\maketitle

\section{Introduction}

In a system of non-interacting electrons disorder can induce
transitions between metallic (delocalized) and insulating
(localized) phases \cite{Anderson1958}. The scaling theory of
localization \cite{Abrahams1979} predicts that in two
dimensions (2D), all single particle electronic states are
typically localized by arbitrarily weak disorder and
consequently, there are no metal-insulator transitions for
non-interacting electrons (except in the situations discussed
below). This scaling idea can be substantiated quantitatively
through a non-linear sigma model (NLSM) formulation of the
problem using either the replica\cite{Wegner1979} or
supersymmetry\cite{Efetov1983a} method. The NLSM approach has
been very successful in providing a quantitative understanding
of the metal-insulator critical point in dimensions greater
than two (for a review, see [\onlinecite{Efetov1997},
\onlinecite{Mirlin2000}]).

There are two well known exceptions to the above discussion.
One is the 2D integer quantum Hall (IQH) plateau transition and
the other is the metal-insulator transition in 2D systems with
spin-orbit scattering (the so-called `symplectic symmetry
class'). These possess phases of both localized as well as
extended states in 2D. On one hand, both these transitions can
be formulated as NLSM field
theories\cite{Levine1983,Hikami1980} (containing an additional,
topological term in the IQH case). But, on the other hand,
these field theories cannot be used to perform analytical (or
even perturbative) calculations of critical exponents or other
(universal) physical quantities at the transition. This is
because the long-distance physics relevant for the properties
at the transition is governed by the strong-coupling regime of
these theories, which is in general not tractable. Indeed, such
a calculation would require a non-perturbative approach to
these problems. Some progress on these types of problems has
been possible in recent years due to the appearance of models
which are analytically tractable even at strong coupling and
which exhibit similar localization-delocalization (LD)
transitions, but belong to symmetry (and universality) classes
different from the ones discussed above. Indeed, seminal work
due to Zirnbauer, and Altland and
Zirnbauer\cite{Zirnbauer1996,Altland1997}, which appeared a
little more than a decade ago, has identified a total of ten
symmetry classes, describing in principle all possible
behaviors of non-interacting quantum mechanical particles
subject to random disorder potentials. Examples of such
analytically tractable 2D localization-delocalization (LD)
transitions include 2D Dirac fermions subject to random Abelian
\cite{Ludwig1994} and non-Abelian gauge potentials
\cite{Nersesyan1994,Mudry1996}, the spin quantum Hall (SQH)
plateau transition
\cite{Senthil1998,Kagalovsky1999,Senthil1999b,Gruzberg1999}
(reviewed briefly in Section \ref{physicsC} below), and
problems with a special so-called `sublattice symmetry'
\cite{Guruswamy2000}.

LD transitions are different in several respects from
continuous phase transitions occurring in non-random systems,
such as the ordering transition in a clean ferromagnet. One
difference is the absence of an obvious order parameter in many
cases. The disorder-averaged density of states which would be
expected to play the role of an order parameter, is not
critical\cite{Wegner1979} at many conventional LD transitions,
including those mentioned at the beginning of this Section.
Another very important difference is that the wave functions at
the transition exhibit what is known as multifractal scaling
behavior
\cite{Castellani1986,Janssen1994,Huckestein1995,Mirlin2000}.
This means that the disorder average of the $q$-th power of the
(square of) the critical wave function amplitude scales with
the system size $L$ with a non-trivial exponent that has a
non-linear dependence on the power $q$. In field-theoretic
language, multifractality can be understood as the existence of
a certain infinite set of fields in the theory which exhibit
infinitely many independent anomalous dimensions with certain
very special (`convexity') properties \cite{Duplantier1991}. In
contrast, analogous infinite sets of fields cannot exists in
`conventional', non-random phase transitions. (In 2D, for
example, such clean phase transitions are known to possess only
a finite number of independent anomalous
dimensions\cite{Belavin1984}.) On one hand, the multifractal
scaling properties of wave functions at the 2D IQH plateau
transition and at the 2D symplectic class metal-insulator
transition have been calculated numerically to high precision
\cite{Evers2001,Mildenberger2007,Obuse2007}. On the other hand,
the entire multifractal spectrum has been obtained analytically
for Dirac fermions in random gauge
potentials\cite{Ludwig1994,Mudry1996,Castillo1997,Caux1998}
while the first few multifractal wave function moments have
been calculated analytically for the spin quantum Hall
transition\cite{Evers2003,Mirlin2003}.

The aforementioned properties, while being important characteristics of the LD critical points, do not exhaust all
their important universal physics. It is well known that even for non-random, conventional phase transitions, the
geometry of the system has a crucial effect on the critical behavior \cite{Binder1983}. Critical exponents as well as
correlation functions depend on the overall geometry and the location where these quantities are measured. In
particular, correlation functions of quantities located near the boundary of the system exhibit in general different
scaling behavior\cite{DiehlDietrich1,DiehlDietrich2,Diehl} as compared to bulk correlation functions (and this depends
in general also on the imposed boundary conditions). These aspects can be studied in great detail especially in the
case of critical 2D systems which are conformally invariant \cite{Cardy1984}. These ideas thus emphasize the importance
of boundary effects in LD transitions, particularly in 2D where these transitions are believed to be in general
described by conformally invariant fixed points (see also Ref.~\onlinecite{Obuse2007}). The SQH transition, being
analytically tractable, provides us with a playground where we can understand the effect of both multifractality and
boundary criticality at LD transitions. In the process we are led to a more general framework where one considers
multifractal behavior of wave functions near boundaries, giving rise to the notion of `{\it surface (boundary)
multifractality}'. The power of conformal invariance enables us to extend this idea to more complicated geometrical
settings. In the present paper we derive several exact exponents characterizing these properties for the SQH
transition.

To summarize our main results, we note that within a
second-quantized formulation of the conventional
Chalker-Coddington network model for the SQH
transition\cite{Kagalovsky1999,Gruzberg1999}, a reflecting
boundary (where the system simply ends) preserves the full
sl$(2|1)$ supersymmmetry present in the disorder-averaged bulk
problem. After the mapping\cite{Gruzberg1999} to bond
percolation this amounts to a reflecting boundary condition for
the perimeters (hulls) of the percolation clusters. Here we
express correlation functions characterizing wave function
multifractality in terms of the generators of the underlying
sl$(2|1)$ supersymmmetry algebra. Using the representation
theory of this superalgebra together with the percolation
mapping we are able to convert the correlation functions to
classical percolation probabilities. Using previously known
scaling exponents for percolation we are able to calculate two
non-trivial multifractal exponents, conventionally denoted by
$\Delta_2$ and $\Delta_3$, both on the boundary and in the
bulk, within the supersymmetry approach developed in Ref.~\onlinecite{Gruzberg1999}. (The values for $\Delta_2$ and
$\Delta_3$, \emph{in the bulk} had been obtained in Refs.~\onlinecite{Evers2003} and \onlinecite{Mirlin2003} using methods
very different from the supersymmetry methods used in the
present paper.) The inability to calculate higher multifractal
moments is explained within our (supersymmetry) formulation.
The scaling behavior of  the averaged local density of states
and point-contact conductance are also found at the boundary.
Using suitable conformal invariance arguments, we are able to
translate the boundary (`surface') exponents to a wedge
geometry and find the corresponding corner exponents (see
Section \ref{Other Geometries} for the various exact values).

The organization of this paper is as follows. In Section
\ref{physicsC} we briefly present the physics of the SQH
effect. In Section \ref{SQH network model}, we discuss the
supersymmetry (SUSY) formulation of the Chalker-Coddington
network model for the SQH transition in the presence of
boundaries. In Section \ref{Multifractal Exponents}, we
elucidate how supersymmetry in the SQH problem can be used to
study (low-order) multifractal wave function moments, both near
boundaries and in the bulk.  (As mentioned above, exponents of
low-order multifractal moments in the {\it bulk} have been
obtained previously in Ref.~\onlinecite{Mirlin2003} using
very different methods.) We also explain, using the
supersymmetry approach, following Ref.
\onlinecite{Gruzberg1999}, the technical reasons why this
computation can be done only for low multifractal moments (up
to the third power of the wave function intensity) and not for
the whole multifractal spectrum. The reasons appear to be
different from, and complementary to the ones discussed in Ref.
\onlinecite{Mirlin2003}. In Section \ref{therm & trans
exponents} we discuss the scaling behavior of the local density
of states and of the point contact conductance near a boundary.
In Section \ref{Other Geometries}, we extend our discussion to
more complicated geometries. In the final Section, we conclude
by discussing certain implications of our results and a number
of remaining open issues. Some of the results derived in the
present paper were announced in the Letter Ref.~\onlinecite{Subramaniam2006}.

\section{Physics of the spin quantum Hall transition }
\label{physicsC}

The spin quantum Hall (SQH) transition was first studied,
numerically, in Ref.~\onlinecite{Kagalovsky1999}. A simple
physical picture of the SQH effect was given in Ref.~\onlinecite{Senthil1999b}. In this Section we briefly review,
for completeness, the basic physics of the SQH effect.

Let us consider a lattice version of the BCS Hamiltonian
describing a singlet superconductor (classes C or CI in the
notation of Refs.~\onlinecite{Zirnbauer1996} and
\onlinecite{Altland1997}):
\begin{align}
\label{BCSHam} {\cal H} = \sum_{ij} \biggl( t_{ij}
\sum_\alpha \cd_{i\alpha} \cph_{j\alpha} +
\Delta_{ij}^{\vphantom{*}} \cd_{i\up} \cd_{j\down} + \Delta^*_{ij}
\cph_{j\down} \cph_{i\up} \biggr).
\end{align}
The first term here describes (in a second-quantized language)
hopping of electrons of either spin $\alpha = \up, \down$
between lattice sites $i$ and $j$, as well as possible on-site
potentials (for $i=j$). The other two terms describe BCS
singlet pairing. Provided that the gap function satisfies
$\Delta_{ij} = \Delta_{ji}$, the Hamiltonian (\ref{BCSHam}) is
SU(2) invariant and commutes with the three generators of the
global SU(2) spin rotations (total spin):
\begin{align}
{\vec S} = \frac{\hbar}{2} \sum_i \cd_{i\al}
{\vec \sigma}_{\al\be}^{\vphantom{\dagger}} \cph_{i\be}.
\label{gens}
\end{align}

Being the Hamiltonian of a superconductor, Eq. (\ref{BCSHam})
does not conserve the particle number, or charge. Physically
this is due to the existence of the pair condensate which may
exchange electrons into holes in processes similar to Andreev
reflection. Not being a conserved quantum number, the
quasiparticle charge cannot be transported by diffusion.
Consequently, there is no notion of electrical (charge-)
conductivity. However, the spin of each quasiparticle is
conserved by the Hamiltonian \eqref{BCSHam}, and one may define
the spin conductivity as the linear response coefficient
between the (say) $z$-component of the spin current and the
gradient of a Zeeman field along the $z$ direction:
\begin{align}
j_i^z = - \sigma_{ij}^{\text{s}} g \mu_B \partial_j B^z.
\end{align}
Here $g$ is the gyromagnetic ratio, and $\mu_B$ is the Bohr
magneton.

When the gap function $\Delta$ is complex, time-reversal symmetry is broken, and then the Hall (transverse) spin
conductivity  $\sigma_{xy}^{\text{s}}$ may be non-vanishing. It is convenient to perform a particle-hole transformation
on the down-spin particles:
\begin{align}
d_{i\up} &= c_{i\up}, & \dph_{i\down} &= \cd_{i\down}.
\end{align}
Under this transformation the $z$ component of the generator of
spin rotations (\ref{gens}) becomes simply the total number of
$d$-particles, $\tfrac{\hbar}{2}\sum_{i,\al} \dd_{i\al}
\dph_{i\al}$. Thus the transformation interchanges the role of
particle number and $z$ component of spin, with the result that
the transformed Hamiltonian conserves the number of $d$
particles:
\begin{align}
{\cal H} = \sum_{ij} \dd_{i\al} H_{ij,\al\be} \dph_{j\be}.
\label{dHam}
\end{align}
Here $H_{ij}$ is the so called Bogoliubov-de Gennes (BdG)
Hamiltonian, a $2 \times 2$ matrix in the spin space:
\begin{align}
\label{BdGHam} H_{ij} =
\begin{pmatrix}
t_{ij} & \Delta_{ij} \\
\Delta_{ij}^* & -t_{ij}^* \\
\end{pmatrix}.
\end{align}
Since the $d$-particle number is conserved we can use a
single-particle description. The external non-random Zeeman
field $B^z$ in the $z$ direction, in this alternative
description maps onto a simple shift $\ep$ in the Fermi energy
as
\begin{align}
\ep = g \mu_B B^z. \label{EB}
\end{align}
The spin conductivity in the original problem becomes the usual
electrical (charge) conductivity of the $d$ particles. Note
that the role of the elementary charge is now played by
$\hbar/2$, so that the natural quantum of the spin conductivity
is $(\hbar/2)^2/2\pi\hbar = \hbar/8\pi$. In what follows we use
units in which $\hbar = 1$.

Invariance of the Hamiltonian in Eq. \eqref{BCSHam} under the
global SU(2) symmetry implies the following property of the
matrix $H_{ij}$ in Eq. \eqref{BdGHam},
\begin{align}
\sigma_y H_{ij} \sigma_y = - H_{ij}^*,
\label{BdGsymm}
\end{align}
which has important consequences.  Firstly, it implies that the
eigenvalues of $H_{ij}$ always appear in pairs as $(\ep,
-\ep)$. In other words, the single particle energy spectrum is
symmetric about $\ep=0$. The ground state of the problem is
obtained by filling up all the negative energy ($d$-) single
particle states. The positive energy ($d$-) particle and hole
excitations on top of this ground state are doublets under the
global SU(2) symmetry. A Zeeman term, or non-zero energy shift
(see Eq. (\ref{EB})), breaks the global SU(2) symmetry and
splits the degeneracy between the above-mentioned ($d$-)
particle and hole states. Secondly, Eq. (\ref{BdGsymm}) implies
a relation between the retarded and advanced Green's functions
of the BdG Hamiltonian or the corresponding network model, (see
Eq. \eqref{advanced and retarded relation} below).

A clean (that is non-random) $d_{x^2 - y^2} + id_{xy}$ -wave
superconductor (which breaks time-reversal symmetry) can, for
example, be obtained on the lattice with a gap function
$\Delta_{ij}$ whose Fourier transform is (square lattice
``$d$-wave''):
\begin{align}
\Delta_k = \Delta_0 (\cos k_x - \cos k_y) - i\Delta_{xy} \sin
k_x \sin k_y,
\label{dwave}
\end{align}
It is easy to see\cite{Senthil1999b} that this represents at
low energies a (2+1)-dimensional Dirac fermion of mass
$\Delta_{xy}$, leading\cite{Ludwig1994} to a quantized spin
Hall conductivity (in units of $\hbar/8\pi$):
\begin{align}
\sigma_{xy}^{\text{s}} = 2 \, \text{sgn} \, \Delta_{xy}.
\end{align}
This is consistent with the existence of two spin-current
carrying states at the edge of a system with a boundary (say,
to vacuum).

These edge states, being chiral, survive the addition of weak
disorder, so the quantization of $\sigma_{xy}^{\text{s}}$
persists in a disordered $d$-wave superconductor with broken
time reversal invariance, at least for weak disorder.  In
general, there can be two possible localized phases
--- spin insulator, and SQH phase, --- distinguished
topologically by the quantized value of the spin Hall
conductivity. A transition between them, the generic SQH
transition, is a localization-delocalization transition similar
to the IQH plateau transition, where $\sigma_{xy}^{\text{s}}$
jumps by two units. It is a quantum percolation transition
\cite{Trugman1983} of the edge states forming\cite{Ludwig1994}
at the interfaces between two topologically distinct regions.

There is one essential difference between the SQH and the IQH
case. In the latter case, the mean single-particle density of
states is non-vanishing on either side and at the  transition.
In the SQH case, on the other hand, the density of states of
$d$-particles, $\rho(\ep)$, vanishes at zero energy, $\ep=0$,
in both localized phases\cite{Senthil1999a} as $\ep^2$; right
at the transition, that is at $\ep=0$, it vanishes (as might
have been expected) with a non-trivial exponent:
\begin{align}
\rho(\ep) \sim |\ep|^\al.
\end{align}
The exponent $\alpha$ was calculated exactly in Ref.~\onlinecite{Gruzberg1999} to be  $1/7$. In Section \ref{therm &
trans exponents} below we will present a derivation of this result, as well as its analogue for the local density of
states at a boundary of the SQH system.

\section{SQH Network Model}
\label{SQH network model}

\subsection{Description of the network}

Network models have been very convenient for both numerical and
analytical work on various disordered non-interacting fermion
problems. The best known such model is the Chalker-Coddington
(CC) network model \cite{Chalker1988} for the integer quantum
Hall plateau transition.

Similar network models can be constructed for other
localization problems, including the chiral metal
\cite{Chalker1995,Gruzberg1997}, the SQH transition
\cite{Kagalovsky1999,Senthil1999b,Gruzberg1999}, the random
bond Ising model and the thermal quantum Hall
effect\cite{Gruzberg2001,Read2001b,Merz2002,Chalker2002}.

\begin{figure}
\begin{center}
\includegraphics[width=0.9\columnwidth]{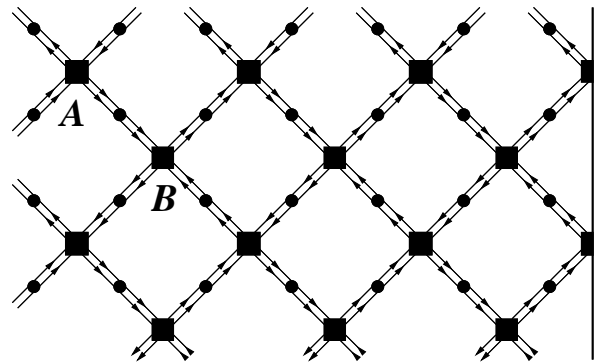}
\caption{SQH network with a vertical reflecting boundary. The black squares represent the non-random node scattering
matrices [Eq. \eqref{scattering matrix}]. The black circles on the links represent the random SU(2) scattering matrices.}
\label{network}
\end{center}
\end{figure}

The SQH network consists of a lattice of directed links and two types of nodes, $A$ and $B$, forming a square lattice
(see Fig. \ref{network}) on which spin-$1/2$ particles at energy $\ep = 0$ can propagate. Uni-directional propagation
through each link is represented by a random SU$(2)$ matrix. As in the case of CC network, to study the critical
behavior at the SQH transition it is sufficient to introduce disorder only for propagation along the links, while all
the nodes can be taken to have the same (non-random) scattering matrices. The node scattering matrices are diagonal in
the spin indices: $S_S=S_{S\uparrow} \otimes S_{S\downarrow}$,
\begin{align}\label{scattering matrix}
S_{S\sigma}=
\begin{pmatrix}
(1-t^2_{S\sigma})^{1/2} & t_{S\sigma} \\
-t_{S\sigma} & (1-t^2_{S\sigma})^{1/2} \\
\end{pmatrix},
\end{align}
where $S=A,B$ labels whether the node is on the $A$ or the $B$
sublattice, and $\sigma = \uparrow,\downarrow$ labels the
spin-index of the propagating particle. Apart from the case of
boundary nodes (which will be treated in detail later), the
remaining network is isotropic (invariant under 90 degree
rotation of the lattice) when the scattering amplitudes on the
two sublattice nodes are related by $t^2_{A\sigma} +
t^2_{B\sigma} = 1$.
The critical point of the isotropic network is located at
$t_{A\sigma} = t_{B\sigma} = 1/\sqrt2$. Varying $t_{S\sigma}$
while keeping $t^2_{A\sigma} + t^2_{B\sigma} = 1$ and
$t_{S\uparrow}=t_{S \downarrow}$ drives the system between a
spin insulator and a SQH state and the spin Hall conductance
jumps from $0$ to $2$. Taking $t_{S\uparrow} \ne t_{S
\downarrow}$ breaks the global SU$(2)$ symmetry and splits the
transition into two ordinary IQH transitions each in the
unitary symmetry class (investigated, for example, in Ref.~\onlinecite{Chalker1988}). In this paper we consider only the
spin-rotation invariant case with $t_{S\up} = t_{S\down} =
t_S$.

\subsection{Green's functions and symmetries}
\label{Greens functions and symmetries}

Network models can be studied using either a first quantized or
a second quantized formalism. The first quantized method is
adopted in Refs.~\onlinecite{Beamond2002} and
\onlinecite{Mirlin2003}. It is also very useful for numerical
work. Here we use it to derive certain symmetry relations. The
rest of the paper will employ the  second quantized formalism
(see next Section).

All physical quantities of interest in our problem such as wave
function correlators and conductance can be expressed in terms
of the Green's functions. Usually, the latter are represented
in continuous time notation, but the network models use the
discrete time analogs. In particular, network models use a
single step time-evolution operator $\mathcal{U}$ which acts on
the single particle wave function $\Psi(r, t)$ at discrete time
$t$ to give the wave function at time $t+1$, where $r$ denotes
a link of the network. For a network with $N$ links,
$\mathcal{U}$ is (due to the spin-index) a $2N \times 2N$
matrix, which represents a finite-time version of the
infinitesimal time-evolution operator. Thus $\mathcal{U} =$
$\exp \{ i {\cal H} \}$ where the Hamiltonian ${\cal H}$ (which
describes the time evolution of the edge states separating
`puddles' of topologically distinct regions) has the same
symmetry properties as the underlying BdG Hamiltonian in Eq.
(\ref{BdGHam}). Therefore, due to the property in Eq.
(\ref{BdGsymm}), $\mathcal{U}$ is a (unitary) symplectic matrix
and as such satisfies the condition
\begin{align}
\label{Usp2N} \mathcal U^{-1} =\sigma_y \mathcal U^T \sigma_y.
\end{align}
where $\sigma_y$ is the conventional Pauli matrix.

Now we can write the retarded (advanced) Green's function $
G_R(\1, \2)$ ($G_A(\1, \2)$) as the resolvent of the operator
$\mathcal U$:
\begin{align}
\bigl[G_R(\1, \2)\bigr]_{\al \be} &= \Big \langle \1,\al \Big|
\frac{1}{1 - z \mathcal U} \Big| \2,\be \Big \rangle,
\label{GR} \\
\bigl[G_A(\1, \2)\bigr]_{\al \be} &= \Big \langle \1,\al \Big|
\frac{1}{1 - z^{-1} \mathcal U} \Big| \2,\be
\Big\rangle.
\label{Greens2}
\end{align}
Here $z=e^{i(\ep + i \ga)}$ where $\ep$ is the energy and
$\gamma$ is a finite level broadening. Since the SQH transition
of interest to us occurs only at zero energy, we will set
$\ep=0$ and hence $z = e^{-\ga}$ from here on. In the above
equation, $ \1, \2 $ and $\al, \be$ denote the network link and
the spin index, respectively.

Making use of Eq. (\ref{Usp2N}) in Eqs. (\ref{GR},
\ref{Greens2}) leads to the following relationship between
advanced and retarded Green's functions
\begin{align}
\bigl[ G_A(\1, \2)\bigr]_{\alpha\pr \alpha} \z\z -
\de_{\alpha\pr \alpha} \delta_{\1, \2} = \epsilon_{\alpha\be} \
\bigl[ G_R(\2, \1)\bigr]_{\be \be\pr} \epsilon_{\be\pr \alpha\pr},
\label{advanced and retarded relation}
\end{align}
where $\ep_{\al \be}$ is the antisymmetric Levi-Civita tensor.
The above equation will turn out to be crucial for our
calculations since it implies that we need not introduce
separate operators for the advanced sector Green's functions
but can work solely with retarded ones (as we will see below).

\subsection{Second-quantized description in the bulk}
\label{bulk network SUSY }

All network models can also be studied using the
second-quantized SUSY technique\cite{Read????,Gruzberg1997}.

The second-quantization of the SQH network is decribed in Ref.~\onlinecite{Gruzberg1999} and we review the main steps here.
The basic idea is quite similar to the transfer matrix
formulation of various 2D statistical mechanics problems such
as the Ising model (for a review, see Ref.~\onlinecite{Baxter1982}). The presence of disorder introduces
a number of additional features which are sketched below.

All physical quantities in the network model including
conductance, wave function amplitudes etc. may be expressed in
first quantization in terms of sums over paths on the network.
Such a sum may be written in second-quantized SUSY language as
a correlation function, $\langle\ldots\rangle\equiv{\rm STr}
\big[ \,\ldots (U_B U_A)^{L_T} \big]$ where the supertrace
`${\rm STr}$' contains the row-to-row transfer matrices $U_A$
and $U_B$ and the ellipsis $\ldots$ stands for operators that
are inserted at the beginning and the ends of paths and
correspond physically to density, current, etc. $L_T$ is the
number of $A$ nodes (or $B$ nodes) along the vertical
direction, interpreted as discrete (imaginary) time. (The
operator $\left(U_B U_A\right)^{L_T}$ was denoted by $U$ in
Ref.~\onlinecite{Gruzberg1999}). The supertrace STr
implements periodic boundary condition along the vertical
direction. Let there be $L_X$ nodes in each row (the
`horizontal', or $X$ direction). The operator $U_A$ (resp.
$U_B$) is formed by multiplying all the $L_X$ transfer matrices
at $A$ (resp. $B$) nodes in a given row. These matrices act in
a tensor product of bosonic and fermionic Fock spaces defined
for each vertical column of links.
The quantum state in the Fock spaces in each horizontal row of
links should be thought of as resulting from the quantum state
of the previous (in discrete imaginary time) row upon action of
either $U_A$ or $U_B$, within a single time step. The presence
of a fermion or boson on a link represents an element of a path
traversing that link \cite{Read????,Gruzberg1997}. Both bosons
and fermions are needed to ensure the cancelation of
contributions from closed loops (this ensures that {\rm
STr}\,$\left(U_B U_A\right)^{L_T} = 1$).

Usually one needs two types of bosons and fermions, `retarded'
and `advanced', to be able to obtain two-particle properties
(that is averages of products of retarded and advanced Green's
functions) relevant for the calculation of transport
properties. However, the symmetry relation in Eq.
\eqref{advanced and retarded relation} relates retarded and
advanced Green's functions. Hence, as it turns out, for the
computation of low  enough moments of, say, retarded Green's
functions,  we need only one fermion and one boson per spin
direction per link of a row. (This will be discussed in detail
below.) We denote them by $f_\sigma$, $b_\sigma$ for the links
going up (up links), and $\fb_\sigma$, $\bb_\sigma$ for the
down links. On the up links, $f_\sigma$, $b_\sigma$ satisfy
canonical fermion and boson commutation relations,
respectively. But to ensure the cancellation of closed loops, we must either take the fermions on the down links to satisfy the modified anti-commutation relations:  $\{\bar{f}_{\sigma},\bar{f}^{\dagger}_{\sigma}\}= - \delta_{\sigma, \sigma^{\prime}}$ (while the bosons on the down links, $\bar{b}_{\sigma}$ and $\bar{b}^{\dagger}_{\sigma}$ satisfy the canonical commutation relations), or, alternatively, we take the bosons on the downlinks to satisfy the modified commutation relations: $[\bar{b}_{\sigma},\bar{b}^{\dagger}_{\sigma}]= - \delta_{\sigma, \sigma^{\prime}}$ (while the fermions on the down links, $\bar{f}_{\sigma}$ and $\bar{f}^{\dagger}_{\sigma}$ satisfy the canonical anti-commutation relations). Each node transfer matrix can be written in
terms of these bosons and fermions. For any given disorder realization, this node transfer matrix turns out to commute with all the generators of the sl$(2|1)$
superalgebra, as shown in Ref.~\onlinecite{Gruzberg1999} (see Appendix A for a summary of the relevant superalgebra representation theory). Performing the average over disorder (independently on each link) projects \cite{Gruzberg1999} the Fock space of bosons and fermions into the fundamental (dual-fundamental) 3 dimensional representation of sl$(2|1)$ on up links (down links). These are precisely the states which are singlets under the random (`gauge') SU(2) on the links.

\subsection{Network boundary}
\label{Network boundary}

The extension of  the above formalism to the case of a network
with a reflecting boundary in the vertical direction is
straightforward. Relegating  some technical details to Appendix
\ref{Boundary SUSY},  the reflecting boundary is seen to
preserve the full sl$(2|1)$ SUSY present in the bulk. Using the
fully intact supersymmetry one can then  retain the mapping to
the perimeters of percolation clusters (we refer to them as
hulls henceforth) obtained  in Ref.
\onlinecite{Gruzberg1999}, but now the percolation hulls are
confined to the half plane in 2D with a reflecting boundary
(see Fig. [\ref{network}]).

We will consider below correlation functions of operators in the network model in the vicinity of the SQH transition.
When using continuum notation one needs to recall\cite{DiehlDietrich1,DiehlDietrich2,Diehl} that, in general, operators
acquire additional singularities when approaching a boundary; they may vanish or diverge.  By saying that points lie on
the boundary, we imply that these points are not literally located {\it at} the boundary, but rather that the distance
of the points from the boundary is much smaller than the distance of these points from each other, and much smaller
than the system size.  The same general reasoning will apply when we consider multifractality of wave functions at the
boundary. Let us recall that in the literature of surface transitions in ordinary magnets, various kinds of cases which
are often referred to as `ordinary, `extraordinary' and `special' boundary transitons are discussed, referring to the
respective ordering of bulk and boundary. The case of the reflecting boundary condition that we consider here is
analogous to the so-called `ordinary' surface transition where the boundary and the bulk undergo the LD transition
simultaneously.

We will need the bulk and  boundary scaling dimensions of the
1-hull operator of critical percolation in various calculations
below. This has been derived using a variety of techniques in
the literature \cite{Cardy1984, Saleur1987, Read2007} and is
found to be $x_s = 1/3$ for the boundary 1-hull operator. The
corresponding bulk field has scaling dimension $x_b = 1/4$.

\section{Multifractality using Supersymmetry}
\label{Multifractal Exponents}

\subsection{Wave function correlators}
\label{wavefncorr.}

In the context of LD transitions, multifractality  manifests
itself through the anomalous scaling with system size of the
moments of the square of the wave function amplitude,
$|\psi(r)|^2$ (wave function `\emph{intensity}') at criticality
(see, for example, Refs.~\onlinecite{Mirlin2000} and
\onlinecite{Janssen1994}).  The wave function moments
conventionally appear in the form of the so-called  averaged
generalized inverse participation ratios (IPR) $P_q$ (the
overbar denotes disorder average)
\begin{align}
P_q &= \int_{{\cal M}_x} d^{D_x}r_x   \, \, \ov{\mid \z \psi(r_x)\z \mid^{2q}} \propto L^{D_x} \ov{\mid \z \psi(r_x)\z
\mid^{2q}}, \label{ipr formula}
\end{align}
whose scaling  with linear system size $L$  at a critical point
defines the critical exponents $\tau^x_q$ and $\Delta^x_q$ (often
referred to as `anomalous dimension'),
\begin{align}
\label{DEFtau-q}
P_q &\propto L^{-\tau^x_q}, & \tau^x_q &= d q + \Delta^x_q + q \mu_x - D_x,
\end{align}
Here, $D_x$ is the dimension of the region ${\cal M}_x$ over which the integration is performed in (\ref{ipr formula}).
In particular, we will be interested in the cases of bulk ($x=b$), boundary ($x=s$) and corner ($x=c$) where $D_x$
takes on, respectively, the values $d$, $d-1$ and $d-2$. The corresponding anomalous dimensions will be denoted by
$\Delta^b_q, \Delta^s_q$ and $\Delta^c_q$. One notes that $\Delta^b_1 = 0$ due to the fact that wave functions are
normalized to unity with respect to the whole system (implying $P_{q=1} =1$ and $\mu_b = 0$). There is no such general
constraint for $\Delta^s_1$ and $\Delta^c_1$. However, we adopt the convention of setting them equal to zero and the
corresponding contribution to $\tau^x_1$ is reflected in $\mu_s$ and $\mu_c$  respectively. With this choice, a
nonvanishing exponent $\mu_x$ characterizes a non-trivial scaling behavior of the local wave function intensity at the
sample location $r_x$,
\begin{align}\label{mu-x definition}
L^d \ \ov{|\psi(r_x)|^2} \propto 1 / L^{\mu_x}
\end{align}
which is known to occur in certain non-conventional Anderson localization symmetry
classes\cite{Zirnbauer1996,Altland1997}, as was discussed previously in Ref.~\onlinecite{Subramaniam2006}.

The wave function moments have a natural interpretation as operators in a field theory describing the LD transition. As an aside, it is worthwhile emphasizing that the interpretation of multifractal moments in
terms of field theory operators is a very general one and not just restricted to LD transitions (see Ref.
\onlinecite{Duplantier1991} for a discussion of this point). We have defined multifractal exponents in terms of moments
of wave function at a single point in the sample in Eq. \eqref{ipr formula}. In field theory, for calculating the
multifractal exponent $\Delta^x_q$, (restricting $q$ to be an integer,) it is easier to find the scaling of correlation
function of q operators, each corresponding to the wave function intensity $L^2|\psi (r)|^2$. The $q$-point correlations of critical
eigenfunction intensities take on a convenient scaling form when the distances between all points are equal to the same
scale $r$, that is $\mid \z r_i - r_j \z \mid \sim r \,$ and it follows that
\begin{align}
L^{q(d+\mu_x)} \ov{ |\psi (\1)\psi (\2) \ldots \psi (\q)|^2}
\sim {\Big(\dfrac{r}{L}\Big)}^{\Delta^x_q},
\end{align}
where $L$ is the linear system size.

Off criticality (that is, for $\ep \ne 0$ or $\gamma > 0$ in the SQH system), multifractal scaling behavior holds on
length scales much shorter than the localization length \cite{Gruzberg1999} $\xi_{\gamma} \sim \gamma^{-4/7}$, beyond
which the wave function amplitudes are exponentially small \cite{Evers2003,Mirlin2003}. This implies that off
criticality, the above $q$-point correlator should be written as
\begin{align}
L^{q(d+\mu_x)} \ov{ |\psi (\1)\psi (\2) \ldots \psi (\q)|^2}
&\sim {\Big(\dfrac{r}{\xi_{\gamma}}\Big)}^{\Delta^x_q}, & r
& \ll \xi_{\gamma}. \label{n-point-wavefuntion-correlator}
\end{align}
Assuming that all wave functions at a given energy show statistically identical multifractal scaling behavior, one can
write the L.H.S of the above correlator as
\begin{widetext}
\begin{align}
L^{q(d+\mu_x)} \ov{ |\psi (\1)\psi (\2) \ldots \psi (\q)|^2} = \dfrac { \ov{\sum_{i_1,i_2 \ldots i_q} |\psi_{i_1}
(\1)\psi_{i_2} (\2) \ldots \psi_{i_q} (\q)|^2 \de(\ep_1 - \ep_{i_1}) \de(\ep_2 - \ep_{i_2}) \ldots \de(\ep_q -
\ep_{i_q})} } {\ov{\rho({\ep}_1)}\, \ov{\rho({\ep}_2)} \ldots \ov{\rho({\ep}_q)}},
\end{align}
where $i_1, i_2 \ldots i_q$ are additional quantum numbers required, besides position $r$, to label the wave function
(energy and spin in the SQH case). $\rho$ represents the global density of states. Taking ${\ep}_1, {\ep}_2 \ldots
{\ep}_q \sim \ep$ and $\nolinebreak{\mid \z r_i - r_j \z \mid} \sim r \,$, we see that the exponent $\Delta^x_q$ is
given by the scaling behavior of ${\mathcal{\tilde{D}}^x_q}(\1,\2 \ldots \q ; \ep)$ with respect to $r$, where the
function $\mathcal{\tilde{D}}^x_q$ is defined by
\begin{align}
{\mathcal{\tilde{D}}^x_q}(\1,\2 \ldots \q ; \ep) &= \ov{\sum_{i_1,i_2 \ldots i_q} |\psi_{i_1} (\1)\psi_{i_2} (\2)
\ldots \psi_{i_q} (\q)|^2 \de(\ep - \ep_{i_1}) \de(\ep - \ep_{i_2}) \ldots \de(\ep - \ep_{i_q})}. \label{q pt wavefn
correlator}
\end{align}
\end{widetext}
We understand that the points $r_1 \ldots r_q$ are all chosen to lie in the region ${\cal M}_x$ considered in Eq.
\eqref{ipr formula}.

Using the Green's functions defined in Sec. \ref{Greens
functions and symmetries}, we can write the above wave function
correlator (ignoring overall factors of $(2\pi)^{-q}$) as
\begin{align}
{\mathcal{\tilde{D}}^x_q} (\1,\2 \ldots \q ; z) &\propto \ov{\prod_{k=1}^q \text{tr}\left[G_R(r_k,r_k)-
G_A(r_k,r_k)\right]}. \label{2nd q point function}
\end{align}
Here `tr' denotes the trace over the spin indices. As was
discussed in Ref. \onlinecite{Mirlin2003}, the multifractal
exponents can also be obtained  from a different Green's
function product,
\begin{align}
{\mathcal{{D}}^x_q}(\1,\2 \ldots \q ; z) &\propto \ov{ \text{tr} \prod_{k=1}^q \z \left[G_R(r_k,r_{k+1}) -
G_A(r_k,r_{k+1})\right]}, \label{1st q point function}
\end{align}
where $r_{q+1}\equiv r_1$. When $\nolinebreak{\mid \z r_i - r_j
\z \mid} \sim r \,$, this product shows the same scaling
behavior, $r^{\Delta^x_q}$ as the previous one. Since the
calculations of both functions ${\tilde {\cal D}}$ and ${\cal
D}$ are very similar, we have explained in detail only the
calculation of $\tilde{\mathcal{D}}$. By suitably choosing the
correct SU(2) invariant (see next section), we will extract
below the scaling of $\tilde{\mathcal{D}}$. Our results agree
with those previously obtained in Ref.~\onlinecite{Mirlin2003} for the {\it bulk} case, $x=b$, but
we obtain these results using a different method  (namely that
developed in Ref.~\onlinecite{Gruzberg1999}, using
supersymmetry). For the other cases, $x=s$ (boundary) and $x=c$
(corner), our results are entirely new.

\subsection{Calculation of $\mathbf{\Delta^x_2}$}
\label{Label-for-Subsection-Delta-x-Two}

We start with the calculation of the exponent $\Delta^x_q$ for
the case $q=2$. The corresponding Green's function product that
we have to evaluate is (dependence  on $z=$ $=\exp(-\gamma)$,
defined in  Eqs (\ref{GR}, \ref{Greens2}), is understood)
\begin{widetext}
\begin{align}
{\mathcal{\tilde{D}}^x_2}(\1, \2) \propto \ov{ \text{tr} \left[G_R(\1, \1) - G_A(\1, \1)\right] \text{tr}\left[G_R(\2,
\2)-G_A(\2,\2)\right]}.
\end{align}
Using Eq. \eqref{advanced and retarded relation}, we can write
the above relation in terms of retarded functions alone as
\begin{align}
{\mathcal{\tilde{D}}^x_2}(\1, \2) &\propto \ov{\text{tr}\left[G_R(\1, \1) - 1 \right] \text{tr} \left[G_R(\2, \2) - 1
\right]} \nonumber\\ &= \ov{\text{tr} \,G_R(\1, \1) \text{tr}\,G_R(\2, \2)} - \ov{ \text{tr}\,G_R(\1, \1)} - \ov{
\text{tr}\, G_R(\2, \2)} + 1. \label{2nd 2 pt function}
\end{align}
\end{widetext}
We will use the supersymmetry technique\cite{Efetov1983a} to
implement the disorder average and express\cite{Gruzberg1999}
all averaged Green's functions using the second quantized
formalism in terms of generators of the sl$(2|1)$ Lie
superalgebra. (A summary of certain basic elements of the
relevant representation theory of the sl$(2|1)$ superalgebra is
presented in Appendix \ref{sl(2|1) algebra}.)

The retarded Green's functions can be expressed as expectation
values of canonical boson and fermion operators. (From here on,
all Green's functions are understood as retarded unless they
have an explicit subscript $A$). Specifically, we can write the
following Green's functions, in any fixed realization of
disorder, as (the subscripts below refer to both, the spatial
coordinate and the spin index)
\begin{align}
G_{ij} &= \text{STr}\big[\bph_i \bd_j (U_B U_A)^{L_T}\big] =
\langle \bph_i \bd_j \rangle, \nonumber \\
G_{kn} &= \text{STr}\big[\fph_k \fd_n (U_B U_A)^{L_T}\big] =
\langle \fph_k \fd_n \rangle,
\label{boson2} \\
G_{ij} G_{kn} &= \text{STr}\big[\bph_i \bd_j \fph_k \fd_n
(U_B U_A)^{L_T}\big] = \langle \bph_i \bd_j \fph_k
\fd_n \rangle.
\label{fermion2boson2}
\end{align}
(Here, the fact that $\text{STr}\,(U_B U_A)^{L_T}$ = 1 was
used.) Let us take $i \to (\1,\al_1), j \to (\1,\be_1), k \to
(\2,\al_2), n \to (\2,\be_2)$ in the above equations. The LHS
of the last equation thus reads $G^{\al_1}_{\be_1}(\1, \1)
G^{\al_2}_{\be_2}(\2, \2)$. After performing the disorder
average, only SU(2) singlet combinations are non-vanishing.
Hence we contract both sides with the SU(2) invariant
$\de^{\al_1}_{\be_1}\de^{\al_2}_{\be_2}$ giving $ \text{LHS} =
\ov{\text{tr}\,G(\1,\1) \, \text{tr}\,G(\2,\2)}$. The RHS of
Eq. \eqref{fermion2boson2} is (we can raise and lower indices
trivially since the metric is the unit matrix)
\begin{align}
\text{RHS} &= \de^{\al_1}_{\be_1}\de^{\al_2}_{\be_2} \,
\langle b_{\al_1}(\1)b\dg_{\be_1}(\1)
f_{\al_2}(\2)f\dg_{\be_2}(\2) \rangle \nonumber \\
&= \big\langle \bigl(2B(\1)+1\bigr)\bigl(1-2Q_3(\2)\bigr) \big\rangle.
\end{align}
(The expressions for the generators $B$ and $Q_3$ of the Lie
superalgebra, as reviewed in Appendix A, were used.) Thus the
LHS and RHS together imply:
\begin{align}
&\overline{\text{tr}\,G(\1, \1) \text{tr}\,G(\2, \2)} \nn \\
&\quad = \big\langle \bigl(2B(\1) + 1\bigr)
\bigl(1 - 2Q_3(\2)\bigr)\big\rangle. \label{bq3 g to gen}
\end{align}
Similarly, considering Eq. \eqref{boson2}, let $
k\to(\2,\al_2), n\to(\2,\be_2)$ giving $\text{LHS} =
G^{\al_2}_{\be_2}\left(\2,\2\right)$. To do disorder average,
contract with the SU(2) invariant $\de^{\be_2}_{\al_2}$, giving
$\text{LHS} =\overline{\text{tr} \,G(\2,\2)}$. The RHS of Eq.
\eqref{boson2} gives
\begin{align}
\text{RHS} = \de^{\be_2}_{\al_2}\, \langle
f_{\al_2}(\2)f\dg_{\be_2}(\2) \rangle = \big\langle 1 -
2Q_3(\2)\big\rangle.
\end{align}
Hence from LHS and RHS, we obtain
\begin{align}\label{q3 G to gen}
\overline{\text{tr} \,G(\2,\2)}= \big\langle 1-2Q_3(\2)\big\rangle.
\end{align}
Now using Eq. \eqref{bq3 g to gen} and Eq. \eqref{q3 G to gen},
we can write Eq. \eqref{2nd 2 pt function} as
\begin{align}
{\mathcal{\tilde{D}}^x_2}(\1, \2) &\propto -\bigl\langle B(\1)Q_3(\2)\bigr\rangle. \label{2nd 2 pt fn G to gen}
\end{align}
We will now compute this correlator off criticality, at a
finite correlation length $\xi_{\gamma}$, arising from a
non-vanishing `broadening' $\gamma > 0$, that is $z = \exp(-
\gamma) < 1$.

The angular brackets in Eq. \eqref{2nd 2 pt fn G to gen} denote the
supertrace STr taken over the \emph{full} Fock space of
canonical bosons and fermions on each link of the network
before disorder averaging. As explained in Ref.
\onlinecite{Gruzberg1999}, the disorder average projects this
Fock space into the fundamental three-dimensional
representation of sl(2$|$1) on the up links and the
corresponding dual representation on the down links. To sum
only over the three states in this representation, we use the
notation `str' (see Eq. \eqref{2nd 2pt prob}). We note that the
state with odd fermion number in both of these representations has
has negative norm \cite{Gruzberg1999}. The prefix `s' in str
denotes the fact that these negative norm states contribute to
the trace with a negative sign. This is essential to get the
correct overall factor of unity for each loop traversed
\cite{Gruzberg1999}. The action of the node transfer matrix,
$T_{S\sigma}$, on the tensor product of the disorder-averaged
states on an up link and a neighboring down link can be
represented by a linear combination of the only two sl$(2|1)$
invariant operators in this $3 \times 3$ dimensional vector
space. These are the identity operator and the projection
operator onto the singlet state, with weights $1-t^2_{S\sigma}$
and $t^2_{S\sigma}$ respectively. These weights can be
considered as the classical probability of a state turning
left or right at a given node under the action of the node
transfer matrix. When we multiply all such node transfer
matrices together, the partition function can be represented as
a sum over all classical configurations of densely packed loops
(the closed classical paths along which the disorder averaged
states propagate) on a square lattice. Each loop gets an
overall weight which is the product of the probabilities of
turning in a particular direction at each node. The loops can
be interpreted (see Ref.~\onlinecite{Gruzberg1999} for a
diagrammatic perspective) as the external perimeters of a
cluster percolating along the bonds of a square lattice. This
completes the mapping to percolation identified in Ref.~\onlinecite{Gruzberg1999}.

Now, in order to evaluate a disorder averaged correlator such as the one in Eq. \eqref{2nd 2 pt fn G to gen}, consider
the $3$ dimensional representation on the links  (for example, the case where $r_1,r_2$ lie on up links) and all
calculations are done in this representation. A loop passing through a link corresponds to all three states in the
three dimensional sl(2$|$1) representation propagating on that link. Away from criticality, we assign a factor of
$z^{2(B+Q_3)}$ for each such link since the operator $2(B+Q_3)$ counts the number of states propagating on that link.
The product of these factors along a path on the network gives the same  power of $z$ that occurs in the Taylor
expansion of a matrix element of the Green's function in Eq. (\ref{GR}). After multiplying such factors for all links
through which a given loop passes, we also multiply it with the overall weight coming from the classical probability
of the loop turning at each node as mentioned in the previous paragraph. Further, if there are operators inserted at
specific points on the lattice, the only loops contributing to the partition sum are those which are constrained to
pass through these points. Considering the correlator in Eq. \eqref{2nd 2 pt fn G to gen}, there are two different
kinds of loop configurations which contribute to the partition sum: (1) a single loop passes through both points $r_1$
and $r_2$, (2) two different loops pass through each of these two points, $\1$ and $\2$. These two terms are the
probabilistic versions of the usual connected and disconnected parts of any correlation function. Writing the
contribution for each of these types separately and summing over all possible loop configurations together with their
respective weights, we write Eq. \eqref{2nd 2 pt fn G to gen} as
\begin{widetext}
\begin{align}\label{2nd 2pt prob}
&4\langle B(\1)Q_3(\2)\rangle\nonumber\\
& =\sum_{N_{12},N_{21}} \text{str}
\begin{pmatrix}
1 & 0 & 0\\
0 & 2 & 0\\
0 & 0 & 1
\end{pmatrix}
\begin{pmatrix}
1 & 0 & 0\\
0 & z^{2N_{12}} & 0\\
0 & 0 & z^{2N_{12}}
\end{pmatrix}
\begin{pmatrix}
-1 & 0 & 0\\
0 & 0 & 0\\
0 & 0 & 1
\end{pmatrix}
\begin{pmatrix}
1 & 0 & 0\\
0 & z^{2N_{21}} & 0\\
0 & 0 & z^{2N_{21}}
\end{pmatrix}
P(\1, \2; N_{12}, N_{21}) \nonumber\\
& \quad + \sum_{N,N'} \text{str}
\begin{pmatrix}
1 & 0 & 0\\
0 & 2 & 0\\
0 & 0 & 1
\end{pmatrix}
\begin{pmatrix}
1 & 0 & 0\\
0 & z^{2N} & 0\\
0 & 0 & z^{2N}
\end{pmatrix}
\text{str}
\begin{pmatrix}
-1 & 0 & 0\\
0 & 0 & 0\\
0 & 0 & 1\end{pmatrix}
\begin{pmatrix}
1 & 0 & 0\\
0 & z^{2N'} & 0\\
0 & 0 & z^{2N'}
\end{pmatrix}
P_-(\1, \2; N, N') \nonumber\\
& = -\sum_{N} \big[1 - z^{2N}\big] P(\1, \2; N) -
\sum_{N,N'} \big[1 - z^{2N}\big]\big[1 - z^{2N'}\big]
P_-(\1, \2; N, N'),
\end{align}
\end{widetext}
where $P(\1, \2; N)$ is the probability of a single loop of length $N$ passing through both $\1$ and $\2$, and
$P_-(\1, \2; N, N')$ is the probability that in a given percolation configuration the points $\1$ and $\2$ belong to
two different, non-overlapping loops of lengths $N$ and $N'$, respectively. In the above equation, the probability
factor $P(\1, \2; N_{12}, N_{21})$ is for a path of length $N_{12}$ going from $\1$ to $\2$ and then a path of length
$N_{21}$ returning from $\2$ to $\1$. In the next line, we simplified this using the fact that this is the same as a
loop of overall length, $N = N_{12} + N_{21}$ passing through both the points, and setting $P(\1, \2; N) := P(\1, \2;
N_{12}, N_{21})$. The appearance of these classical percolation probabilities in the above formula arises from  the
product of the individual factors for turning left or right at each node of the network model.  Thus, using Eq.
\eqref{2nd 2pt prob}, we can write Eq. \eqref{2nd 2 pt fn G to gen} as:
\begin{align}
\label{2nd 2 pt fn probability} {\mathcal{\tilde{D}}^x_2}(\1, \2) &\propto \sum_{N}
\big[1 - z^{2N}\big] P(\1, \2; N) \nonumber \\
& + \sum_{N,N'} \big[1 - z^{2N}\big] \big[1 - z^{2N'}\big]
P_-(\1, \2; N, N').
\end{align}
We see that the leading order terms turn out to vanish in the critical limit $z \to 1$, as observed in Ref.~\onlinecite{Mirlin2003}. So we need to consider the sub-leading behavior. The percolation probabilities discussed
above are the Laplace transformed versions of the correlation functions of $1$-hull operators in percolation. The
1-hull operator represents the density-of-states operator in the SQH problem \cite{Gruzberg1999}. We know the scaling
dimension of these operators as mentioned in the last paragraph in Sec. \ref{Network boundary}. Under the Laplace
transform, the variable $N$ is the conjugate of the energy $\ep + i\gamma \equiv - i \ln z$ which is represented by the
$1$-hull operator. Now, using the scaling dimensions of the 1-hull operator  to deduce the scaling of the Laplace
transformed correlation functions (see, for example, Ref.~\onlinecite{deGennes1988}),  one finds the following
leading scaling behavior of the above-mentioned probabilities
\begin{align}
P(\1, \2; N) &\sim N^{-2/(2-x_b)} r^{-x_b} \nonumber \\
&\sim N^{-8/7} r^{-1/4}, \\
P_-(\1, \2; N, N') &\sim P(\1; N) P(\2; N') \nonumber \\
&\sim N^{-8/7} N'^{-8/7},
\end{align}
where $P(\1; N)$ is the probability that the point $\1$ belongs to a loop of length $N$. The points $\1$ and $\2$ lie
in the bulk and $r = |r_1 - r_2| \ll \xi_{\ga} \sim \gamma^{-4/7} \sim N^{4/7}$. Similarly, the scaling expressions in
the case where both points lie on the boundary are
\begin{align}
P(\1, \2; N) &\sim N^{-1 - x_s/(2-x_b)} r^{-x_s} \nonumber \\
&\sim N^{-25/21} r^{-1/3},\\
P_-(\1, \2; N, N') &\sim P(\1; N) P(\2; N') \nonumber \\
&\sim N^{-25/21} N'^{-25/21}.
\end{align}
We see that only the first term in Eq. \eqref{2nd 2 pt fn probability} (which is the connected part) gives rise to the
non-analytic $r$ dependence in the limit $r \ll \xi_{\ga}$, which we are interested in computing. Hence the scaling of
${\mathcal{\tilde{D}}^x_2}(\1, \2)$ is given by:
\begin{align}
{\mathcal{\tilde{D}}^x_2}(\1, \2) &\sim \sum_{N}
\big[1 - e^{-2N \ga}\big] P(\1, \2; N) \nn \\
&\sim
\begin{cases}
r^{-1/4} \text { (bulk) },  \\
r^{-1/3} \text { (boundary)}.
\end{cases}
\end{align}
Thus, upon comparison with Eq.
(\ref{n-point-wavefuntion-correlator}), we finally read off the
multifractal exponent for $q=2$ as being $\Delta_2^{b} = -1/4$
for the bulk, and $\Delta_2^{s} = -1/3$ for the boundary. We
will discuss the case of the wave function scaling behavior at
corners at the end (see Sec. \ref{Other Geometries}) as it is a
straightforward extension of the  boundary case upon making use
of conformal invariance. In the following paragraph we
consider the more interesting exponent describing the scaling
of the third moment of the square of the wave functions
amplitude.

\subsection{Calculation of $\mathbf{\Delta_3}$}

The algebraic procedure for calculating the multifractal
exponent $\Delta_q$ for $q=3$ is almost same as that we used
above for $q=2$.  Hence we present only the important steps and
focus on the main results and the interesting differences. We
start with the expression:
\begin{widetext}
\begin{align}
{\mathcal{\tilde{D}}}^x_3(\1, \2, \3) &\propto \ov{ \text{tr} \left[G_R(\1, \1) - G_A(\1, \1)\right]
\text{tr}\left[G_R(\2, \2) - G_A(\2, \2)\right] 
\text{tr}\left[G_R(\3, \3) - G_A(\3, \3)\right]}.
\end{align}
Converting all advanced Green's functions to retarded ones
using Eq. \eqref{advanced and retarded relation}, we can write
the above equation as:
\begin{align}\label{2nd 3 pt function}
{\mathcal{\tilde{D}}^x_3}(\1, \2, \3) &\propto \ov{\text{tr}\left[G(\1, \1) - 1 \right] \text{tr} \left[G(\2, \2) - 1
\right]
\text{tr}\left[G(\3, \3) - 1 \right]} \nonumber\\
&= \ov{ \text{tr}\,G(\1, \1) \, \text{tr}\,G(\2, \2) \, \text{tr}\,G(\3, \3)} - 1
- \sum_{[\1,\2,\3]} \ov{ \text{tr}\,G(\1, \1) \, \text{tr}\,G(\2, \2)} +
\sum_{[\1,\2,\3]} \ov{ \text{tr}\,G(\1, \1)},
\end{align}
\end{widetext}
where $\sum_{[\1,\2,\3]}$ denotes sum over terms with all
cyclic permutations of $(\1,\2,\3)$.

The new piece we have to evaluate here is the product of three
Green's functions. For this consider the following two
identities, obtained by applying Wick's theorem, and valid in
any fixed realization of disorder:
\begin{align} \label{fermion2boson4}
\left[G_{ij} G_{lm} + G_{im} G_{lj}\right] G_{kn}=
\langle b_i b\dg_j b_l b\dg_m f_k f\dg_n \rangle.
\end{align}
Exchanging bosons and fermions in the above equation, we find
\begin{align} \label{fermion4boson2}
\left[G_{ij} G_{lm} - G_{im} G_{lj}\right] G_{kn}=
\langle f_i f\dg_j f_l f\dg_m b_k b\dg_n \rangle.
\end{align}
In Eq. \eqref{fermion2boson4}, let $i \to (\1, \al_1)$, $j \to
(\1, \be_1)$, $l \to (\2, \al_2)$, $m \to (\2, \be_2)$, $k \to
(\3, \al_3)$, $n \to (\3, \be_3)$, yielding
\begin{widetext}
\begin{align}
\text{LHS} &= [G^{\al_1}_{\be_1}(\1, \1) G^{\al_2}_{\be_2}(\2, \2) +
G^{\al_1}_{\be_2}(\1, \2) G^{\al_2}_{\be_1}(\2, \1)]
G^{\al_3}_{\be_3}(\3, \3).
\end{align}
As above, it is convenient to contract with the SU(2) invariant
$\de^{\be_1}_{\al_1}\de^{\be_2}_{\al_2}\de^{\be_3}_{\al_3}$,
after performing the disorder average
\begin{align}
\label{3A} &\overline{\text{tr}\,G(\1,\1) \, \text{tr}\,G(\2,\2) \,
\text{tr}\,G(\3,\3)} + \overline{\text{tr}\,G(\3,\3) \,
\text{tr}\left[G(\1,\2) G(\2,\1)\right]} \nn \\
& = \langle b_{\al}(\1) b\dg_{\al}(\1) b_{\be}(\2) b\dg_{\be}(\2)
f_{\ga}(\3) f\dg_{\ga}(\3) \rangle
= \big\langle \left[1 + 2B(\1)\right] \left[1 + 2B(\2)\right]
\left[1 - 2Q_3(\3)\right]\big\rangle.
\end{align}
We see that the first term on the LHS of Eq. \eqref{3A} is the
Green's function product we want to evaluate in Eq. \eqref{2nd
3 pt function}, while the second term has to be eliminated.
This can be achieved by  performing steps, analogous to those
above, in Eq. \eqref{fermion4boson2}, yielding
 \nolinebreak{
\begin{align}
\label{3B} &\overline{\text{tr}\,G(\1,\1) \, \text{tr}\,G(\2,\2) \,
\text{tr}\,G(\3,\3)} - \overline{\text{tr}\,G(\3,\3) \,
\text{tr}\left[G(\1,\2) G(\2,\1)\right]} \nn \\
& = \langle f_{\al}(\1) f\dg_{\al}(\1) f_{\be}(\2) f\dg_{\be}(\2)
b_{\ga}(\3) b\dg_{\ga}(\3) \rangle
= \big\langle \left[1 - 2Q_3(\1)\right] \left[1 - 2Q_3(\2)\right]
\left[1 + 2B(\3)\right]\big\rangle.
\end{align}
}%
Adding Eq. \eqref{3A} and Eq. \eqref{3B}, we obtain
\begin{align} \label{3}
2 \, \overline{\text{tr}\,G(\1,\1) \, \text{tr}\,G(\2,\2) \, \text{tr}\,G(\3,\3)}
& = \big\langle \left[1 + 2B(\1)\right] \left[1 + 2B(\2)\right]
\left[1 - 2Q_3(\3)\right] \nonumber\\
& \quad + \left[1 - 2Q_3(\1)\right] \left[1 - 2Q_3(\2)\right]
\left[1 + 2B(\3)\right]\big\rangle.
\end{align}
We can now use Eqs. (\ref{bq3 g to gen}, \ref{q3 G to gen},
\ref{3}) to write Eq. \eqref{2nd 3 pt function} as
\begin{align}\label{2nd 3pt fn G to gen}
{\mathcal{\tilde{D}}_3^x}(\1,\2,\3) &\propto \big\langle \left[1 + 2B(\1)\right] \left[1 + 2B(\2)\right] \left[1 -
2Q_3(\3)\right] + \left[1 - 2Q_3(\1)\right] \left[1 - 2Q_3(\2)\right] \left[1 + 2B(\3)\right] \big\rangle
\nonumber\\
& \quad -\sum_{[\1,\2,\3]} \big\langle
\left[1 - 2Q_3(\1)\right] \left[1 + 2B(\2)\right] \big\rangle
+ \sum_{[\1,\2,\3]} \big\langle 1 + 2B(\1) \big\rangle - 1.
\end{align}
\end{widetext}
%
Converting these expressions to percolation probabilities is
exactly analogous to the two point case. Taking the critical
limit $z \to 1$, we obtain
\begin{align}
{\mathcal{\tilde{D}}}^x_3(\1,\2,\3) &\propto P(\1,\2,\3) +  P(\1,\3,\2).
\end{align}
where $P(\1,\2,\3)$ is the probability of a loop of any size
traversing $\1,\2,\3$ in that order. There is no cancelation to
leading order at criticality here, unlike the two point case.
The final scaling of this correlator is simply given by that of
the usual 3-point correlation function of percolation 1-hull
operators at criticality,
\begin{align}
P(\1,\2,\3), P(\1,\3,\2) \sim
\begin{cases}
r^{-3x_b} \sim r^{-3/4} \text{ (bulk)}, \\
r^{-3x_s} \sim r^{-1}\text{ (surface)}.
\end{cases}
\end{align}
Hence the value of multifractal exponent for $q=3$ is
$\Delta_3^{b} = -3/4$ and $\Delta_3^{s} = -1$, for bulk and
surface, respectively.

\subsection{Higher multifractal exponents}

The procedure for calculating $\Delta^x_3$ was very similar to
that of $\Delta^x_2$ (albeit more tedious) and one might ask if
it can be extended  to higher multifractal exponents,
$\Delta^x_q$ with $q >3$. On the other hand however, the fact
that we are able to calculate $\Delta^x_3$ at all using our
formalism is conceptually surprising because usually in
supersymmetric problems, an additional set (=`replica') of
boson and fermion operators is required for each additional
Green's function factor  entering the product in Eq. \eqref{2nd
q point function}. Here we are able to calculate both two and
three-point functions with the same number of  boson and
fermion operators (replicas).

To understand this, we look at the calculation of $\Delta_3$
carefully. We had an extra `unwanted' Green's function product
in Eq. \eqref{3A}. But we were able to  eliminate it by using
an equation similar to Eq. \eqref{3B},  with bosons and
fermions exchanged.  The unwanted Green's function product
canceled between Eq. \eqref{3A} and Eq. \eqref{3B} when added
together, giving us the exact product that we wanted.

Now does such a cancelation go through for higher point
functions? To answer this, we look at the next higher exponent,
$\Delta^x_4$. Here we will have to evaluate a product of four
Green's functions,
$$
\ov{ \text{tr}\,G(\1, \1) \, \text{tr}\,G(\2, \2) \,
\text{tr}\,G(\3, \3) \, \text{tr}\,G(\4, \4)}.
$$
To evaluate this, we will have to use the following identity
obtained from Wick's theorem  (in a fixed realization of
disorder):
\begin{align}\label{fermion4boson4}
&\left[G_{ij} G_{lm} - G_{im} G_{lj}\right]
\left[G_{pq} G_{rs} + G_{ps}G_{rq}\right] \nonumber \\
&= \langle f_i f\dg_j f_l f\dg_m b_p b\dg_q b_r b\dg_s \rangle.
\end{align}
In this equation, let again $i \to (\1,\al_1)$, $j \to
(\1,\be_1)$, $l \to (\2,\al_2)$, $m \to (\2,\be_2)$, $p \to
(\3,\al_3)$, $q \to (\3,\be_3)$, $r \to (\4,\al_4)$, $s \to
(\4,\be_4)$. Then contracting with the SU(2) invariant
$\de^{\be_1}_{\al_1}\de^{\be_2}_{\al_2}\de^{\be_3}_{\al_3}\de^{\be_4}_{\al_4}$
gives us upon averaging the relation:
\begin{widetext}
\begin{align}
\label{4Aa} &\overline{\text{tr}\,G(\1,\1) \, \text{tr}\,G(\2,\2) \,
\text{tr}\,G(\3,\3) \, \text{tr}\,G(\4,\4)} +
\overline{\text{tr}\,G(\1,\1) \, \text{tr}\,G(\2,\2) \,
\text{tr}\left[G(\3,\4) G(\4,\3)\right]} \nonumber\\
&- \overline{\text{tr}\left[G(\1,\2) G(\2,\1)\right]
\text{tr}\,G(\3,\3) \, \text{tr}\,G(\4,\4)} -
\overline{\text{tr} \left[G(\1,\2) G(\2,\1)\right]
\text{tr} \left[G(\3,\4) G(\4,\3)\right]} \nonumber\\
& \quad = \big\langle\left[1 - 2Q_3(\1)\right] \left[1 - 2Q_3(\2)\right]
\left[1 + 2B(\3)\right] \left[1 + 2B(\4)\right]\big\rangle.
\end{align}
There are three unwanted terms on the LHS. Exchanging bosons
and fermions gives on the other hand:
\begin{align}
\label{4Ab}
&\overline{\text{tr}\,G(\1,\1) \, \text{tr}\,G(\2,\2) \,
\text{tr} \,G(\3,\3) \, \text{tr}\,G(\4,\4)} -
\overline{\text{tr}\,G(\1,\1) \, \text{tr}\,G(\2,\2) \,
\text{tr}\left[G(\3,\4) G(\4,\3)\right]}\nonumber\\
&+ \overline{\text{tr}\left[G(\1,\2) G(\2,\1)\right]
\text{tr}\,G(\3,\3) \, \text{tr}\,G(\4,\4)} -
\overline{\text{tr} \left[G(\1,\2) G(\2,\1)\right]
\text{tr}\left[G(\3,\4) G(\4,\3)\right]} \nonumber\\
& \quad = \big\langle\left[1 - 2Q_3(\3)\right]\left[1 - 2Q_3(\4)\right]
\left[1 + 2B(\1)\right]\left[1 + 2B(\2)\right]\big\rangle.
\end{align}
Adding Eq. \eqref{4Aa} and Eq. \eqref{4Ab} eliminates two
unwanted pieces:
\begin{align}
\label{4A} & 2 \, \overline{\text{tr}\,G(\1,\1) \, \text{tr}\,G(\2,\2) \,
\text{tr}\,G(\3,\3) \, \text{tr}\,G(\4,\4)} -
2 \, \overline{\text{tr}\left[G(\1,\2) G(\2,\1)\right]
\text{tr} \left[G(\3,\4) G(\4,\3)\right]} \nonumber\\
&= \big\langle\left[1 + 2B(\1)\right] \left[1 + 2B(\2)\right]
\left[1 - 2Q_3(\3)\right] \left[1 - 2Q_3(\4)\right]\big\rangle + \nn \\
& \quad \big\langle\left[1 - 2Q_3(\1)\right] \left[1 - 2Q_3(\2)\right]
\left[1 + 2B(\3)\right]\left[1 + 2B(\4)\right]\big\rangle.
\end{align}
\end{widetext}
We are still left with one unwanted piece which cannot be
evaluated or canceled with something else. Any other
combination of the generators of the Lie superalgebra  at four
points will generate more terms on applying Wick's theorem and
cannot be canceled out.

Now we see the special feature of the three point calculation.
We had two Green's function products (by applying Wick's
theorem), of which only one was necessary. The supersymmetric
formulation gave us one more equation due to boson-fermion
interchangeability. The unwanted piece canceled between the
fermionic and bosonic equations. This fact does not help us in
higher correlation functions as they have more unwanted pieces.
It is also clear that situation becomes worse for higher
$n$-point functions. Interestingly, the same conclusion was
reached in a very different way in Ref.~\onlinecite{Mirlin2003}.

\section{Local Density of States and Point Contact Conductance}
\label{therm & trans exponents}

Having considered the multifractal calculation in detail, the
calculation of other boundary critical exponents is completely
analogous and we list only the important results. Some of the
bulk exponents have been found in Refs.
\onlinecite{Gruzberg1999}, \onlinecite{Mirlin2003}, and 
\onlinecite{Beamond2002} using a very different technique.
The averaged local density of states (LDOS), summed over the
spin indices, can be written in terms of Green's functions as
\begin{align}
\langle \rho^x(r,\ep) \rangle &=\frac{1}{4\pi} \ov{\text{tr}
\left[G_R(r,r) - G_A(r,r)\right]} \nonumber\\
&=\frac{1}{2\pi} \ov{\text{tr}\,G_R(r,r)- 1}. \label{ldos}
\end{align}
This can again be expressed in terms of the sl$(2|1)$
supersymmetry generators as $(1/2\pi) \bigl\langle 2B(r)
\bigr\rangle$, and this average, following the same steps
presented in Section \ref{Label-for-Subsection-Delta-x-Two},
can be written in terms of percolation probabilities as
\begin{align}
\bigl\langle 2B(r) \bigr\rangle = 1 - \sum_{N} P(r; N) \cos{2N\ep}.
\end{align}
(The same result was also obtained in Ref.
\onlinecite{Beamond2002} using, as mentioned, different
techniques.)
As we have mentioned in the end of Sec. \ref{SQH network
model}, the boundary and the bulk scaling dimensions of the one
hull operator are $x_s = 1/3$ and $x_b = 1/4$. The latter value
implies that the percolation hull has fractal dimension $2 -
x_b = 7/4$, so that $P(r,N) \sim  N^{-8/7}$ for $r$ in bulk.
This yields, according to Eq. \eqref{ldos}, the following
scaling behavior of the LDOS\cite{Gruzberg1999}
\begin{align}
\rho^b(\epsilon)\propto \epsilon^{x_b/(2-x_b)}
=\epsilon^{1/7}.
\end{align}

Note that $2-x_b = 7/4$ is the dynamic critical exponent
governing the scaling of energy with the system size $L$ at SQH
criticality, so that the level spacing at $\epsilon=0$ (and
thus the characteristic energy of critical states) is
$\delta\sim L^{-7/4}$. In our case, when the point $r$ is
located at the boundary we find
\begin{align}
P(r,N)\sim N^{-1-x_s/(2-x_b)} = N^{-25/21},
\end{align}
and the LDOS scaling
\begin{align}
\rho^s(r,\epsilon)\propto \epsilon^{x_s/(2-x_b)} =
\epsilon^{4/21}.
\end{align}
Here we have used the bulk dynamic critical exponent in determining the energy scaling of the boundary LDOS. This is
because we are dealing with an `ordinary surface transition' and here surface (boundary) criticality is driven by the
divergence of the bulk correlation length \cite{Binder1983}. Thus the scaling of LDOS changes between bulk and
boundary. This means, as was mentioned in Section \ref{wavefncorr.}, that the average of the (square of the) wave
function amplitude is suppressed at the boundary, giving rise to a non-vanishing value of $\mu_{x=s} = 1/3-1/4=1/12$
(see Eq. \eqref{mu-x definition}).

A similar procedure is adopted for the calculation of other
boundary exponents (we give a table of all exponents later).
The  boundary diffusion propagator can be written as
\begin{align}
\langle\Pi_{ss}(\1,\2)\rangle = 2\bigl\langle V_-(\1)W_+(\2)
\bigr\rangle,
\end{align}
where $\1$ and $\2$ lie at the boundary. In terms of
percolation probabilities this reads
\begin{align}
\langle\Pi_{ss}(\1,\2)\rangle
= 2\sum_{N} P\left(\1,\2;N\right) z^{2N}.
\end{align}
Taking the limit $z\to 1$ (critical point), gives
\begin{align}
\langle\Pi_{ss}(\1,\2)\rangle = 2\sum_{N} P\left(\1,\2;N\right)
= 2 P(\1, \2).
\end{align}
Here the probability $P(\1, \2)$ for the points $\1$and $\2$ to
be connected by a hull of any length scales at the boundary as
$r^{-2x_s}$ giving
\begin{align}
\langle\Pi_{ss}(\1,\2)\rangle \sim |\1 - \2|^{-2/3}.
\end{align}

Another physical quantity of interest is the boundary point-contact conductance. In a network model setting, this is
defined as the conductance between two boundary links $\1$ and $\2$ which are cut to make it possible to insert and
extract currents from them \cite{Janssen1999}. In the  second quantized formalism, this is equivalent to creating a boson (or alternately a fermion because of the supersymmetry) for each spin direction at one link and destroying the same particle at another link. This translates to the expression
\begin{align}
\langle g_{point}(\1,\2)\rangle \equiv \langle
f^{\dg}_{\uparrow}(\1)f^{\dg}_{\downarrow}(\1)
f_{\downarrow}(\2)f_{\uparrow}(\2) \rangle.
\end{align}
In terms of sl$(2|1)$ generators, this equals $\bigl\langle Q_+(\1) Q_-(\2) \bigr\rangle$. Skipping here the mapping to
percolation probabilities, we find that the point-contact conductance decays exactly in the same way as the diffusion
propagator, and, in particular, scales as $|\1 - \2|^{-2/3}$ at the boundary, at criticality. This result is expected since the correlators  for the diffusion propagator and the point-contact conductance represent different superspin components in the same representation of the superalgebra sl$(2|1)$.

\section{Exponents in Other Geometries}\label{Other Geometries}

There are two distinct ways of extending our discussion of
boundary behavior. The first one is when some of the points lie
on the  boundary while the others lie in the bulk. The other
case is to consider boundaries with more complicated
geometries, the simplest example of which would be a wedge with
opening angle $\theta$. (The  boundary case, considered in the
previous sections corresponds to $\theta = \pi$).

The two-point quantities  are easily computed when one point
$\1$ lies in the bulk and the other point, $\2$ is  at the boundary. These
scale as $r^{-(\eta_b+\eta_{\|})/2}$ where $\eta_b$ and
$\eta_{\|}$ are the usual exponents giving the decay of the two
point function in the bulk and  along the boundary,
respectively. Hence the diffusion propagator and the
point-contact conductance between a point in the bulk and
another  at the boundary, both scale with distance $r$ as
$r^{-7/12}$.

In the case of multifractal exponents, we can calculate the
scaling behavior of correlation functions similar to those in
Eq. \eqref{2nd q point function}. But we should not interpret
these as representing properties of a single multifractal since
multifractality is essentially a single point property. With
this caveat, the value of the quantity analogous to $\Delta_2$
when one point is in the bulk and another is at the boundary is
$-1/4$.
\begin{table}[h]
\begin{tabular}{|c|c|c|c|}
\hline Geometry & \includegraphics[width=0.6 in]{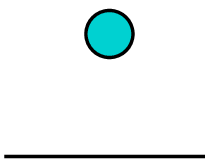}&
\includegraphics [width=0.6 in]{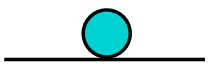}
&\includegraphics[width=0.6 in]{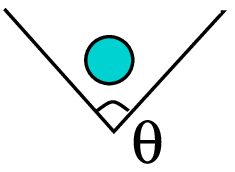}\\
\hline LDOS & $\ep^{1/7}$ & $\ep^{4/21}$& $\ep^{(4/21)(\pi/ \: \theta)}$ \\
\hline $\Delta_2$ & $ -1/4$ & $-1/3$& $-\pi/3 \theta$ \\
\hline $\Delta_3$ & $-3/4$ & $-1$& $-\pi /\theta$ \\
\hline
\end{tabular}
\caption{ One-point exponents in various geometries: bulk, near straight boundaries and near corners. The first line indicates the scaling behavior of the local density of states (LDOS) at the SQH transition. The second and third lines represent the multifractal scaling exponents for the second and third moments of the critical wave function intensity.}
\label{one
point exponents}
\end{table}

\begin{table}[h]
\begin{tabular}{|c|c|c|c|c|}
\hline Geometry & \includegraphics [width=0.5 in]{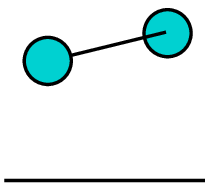}&
\includegraphics [width=0.5 in] {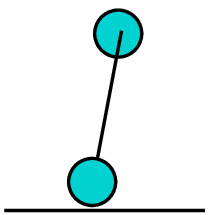}
&\includegraphics [width=0.5 in]{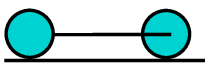}
&\includegraphics [width=0.5 in]{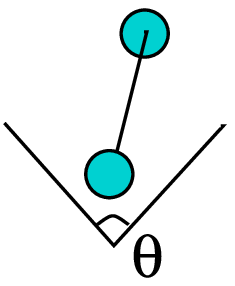}\\
\hline $\Pi(\1,\2)$ & $ r^{-1/2}$ & $r^{-7/12}$& $r^{-2/3}$ &
$r^{-1/4 -(1/3)(\pi / \theta)}$ \\
\hline
\end{tabular}
\caption{Critical scaling behavior of the diffusion propagator, $\Pi(\1,\2)$ in various geometries. The point-contact conductance has an identical scaling behavior as indicated in the text.}
\label{two point exponents}
\end{table}

The next generalization is to analyze the SQH transition in a
wedge geometry. This can be readily carried out using conformal
invariance arguments \cite{Cardy1984}. The conformal
transformation $w = z^{\theta / \pi}$ maps the boundary
geometry in the $z$-plane to a wedge of opening angle $\theta$
in the $w$-plane. From this, it can be deduced that if we
consider a 2-point function with one point lying near the wedge
tip and another deep in the bulk at distance $r$, the two point
function decays as $r^{-\eta_{\theta}}$ where
\begin{align}
\eta_{\theta}=x_b + \frac{\pi}{\theta} x_s.
\end{align}
This enables us to calculate the relevant exponents. In the
density of states calculation, we will have to replace $x_s$ by
$(\pi / \theta) x_s$.

\section{Conclusions and Outlook}\label{Summary}

The central result of this paper is the calculation of boundary
critical and multifractal exponents for the spin quantum Hall
transition in various geometries (see Tables \ref{one point
exponents} and \ref{two point exponents} for a summary of
results). In this paper, we have confined ourselves to the
exact calculation of anomalous multifractal dimensions for low
moment order $q$, using the percolation mapping. In general, in
the theory of multifractality, the information contained in the
set of all the critical exponents $\tau_q^x$ defined in Eq.
(\ref{DEFtau-q}) can also be expressed in terms of the
so-called singularity spectrum $f_x(\alpha)$, its Legendre
transform. We remark that in the presence of a boundary, the
singularity spectrum concept has to be understood more broadly.
This is because unlike in ordinary critical phenomena, the
presence of a  boundary can affect the singularity spectrum
(and $\tau^x_q$) of the entire system,  including bulk and
boundaries, in a significant way, even in the thermodynamic
limit. These issues were first pointed out in Ref.
\onlinecite{Subramaniam2006}, and were analyzed therein, as
well as in Refs. \onlinecite{Obuse2007} and
\onlinecite{Obuse2008}.

The analysis of multifractality in various other geometries
leads to interesting new concepts. Some of these ideas have
been explored for the Anderson transition in $d = 2 +
\epsilon$\cite{Subramaniam2006}, power-law random banded
matrices \cite{Mildenberger2007b} and the 2D symplectic class
transition\cite{Obuse2007, Obuse2008} all of which display the
characteristic multifractal property of LD critical points. A
similar study of Dirac fermions in random gauge fields
\cite{Subramaniam????} and the integer quantum Hall transitions
in two dimensions is expected to further our understanding
of  boundary multifractality and provide important clues
regarding the structure of some of the unknown bulk theories.

We thank N. Read for initial discussions of boundary MF at the
SQH transition and A. D. Mirlin for a previous collaboration on
boundary MF. This work was supported in part by the
Chandrasekhar and Sachs fellowships (ARS), NSF Career award
DMR-0448820, NSF MRSEC DMR-0213745, the Alfred P. Sloan
Foundation and the Research Corporation (IAG), and NSF grant
DMR-0706140 (AWWL).

\appendix
\section{Representations of $\mathbf{\text{sl}(2|1)}$ superalgebra}
\label{sl(2|1) algebra}

The sl$(2|1)$ superalgebra has eight generators, of which four
are bosonic $(B, Q_3, Q_+, Q_-)$ and four are fermionic $(V_+,
V_-, W_+, W_-)$. We use the convention of Ref.~\onlinecite{Scheunert1977} for the generators. These satisfy
the same commutation ([,\,]) and anticommutation (\{, \})
relations as the generators of osp$(2|2)$ superalgebra:
\begin{gather}
[B,Q_3] = [B,Q_\pm] = 0, \nonumber \\
\begin{align}
[B,V_\pm] &= \frac{1}{2} V_\pm, & [B,W_\pm] &= -\frac{1}{2}
W_\pm, \nonumber \\
[Q_3, Q_\pm] &= \pm Q_\pm, & [Q_+, Q_-] &= 2Q_3, \nonumber \\
[Q_3, V_\pm] &= \pm \frac{1}{2}V_\pm,
& [Q_3, W_\pm] &= \pm\frac{1}{2}W_\pm, \nonumber \\
[Q_+, V_-] &= V_+, & [Q_+, W_-] &= W_+, \nonumber \\
[Q_-, V_+] &= V_-, & [Q_-, W_+] &= W_-, \nonumber
\end{align} \\
[Q_+, V_+] = [Q_+, W_+ ] = [Q_-, V_-] = [Q_-,W_- ] = 0,
\nonumber \\
\{V_+,V_-\} = \{W_+,W_-\} = 0, \nonumber \\
\{V_+, W_+\} = Q_+, \quad\,\,\, \{V_+, W_-\} = B-Q_3, \nonumber \\
\{V_-, W_-\} = -Q_-, \quad \{V_-, W_+\} = -B - Q_3.
\end{gather}

An important subalgebra is gl$(1|1)$ formed by the generators $(B, Q_3, V_-, W_+)$. This is the SUSY present in the SQH
network at finite broadening $\gamma$, when $z = e^{-\gamma} < 1$ (see Eqs. (\ref{GR}, \ref{Greens2})), as well as at
an absorbing boundary.

The sl$(2|1)$ algebra has an oscillator realization formed by
bilinear combinations of the fermion and boson operators on
each link that are SU(2) singlets. For the up-links the
oscillator representation is:
\begin{align}
Q_3 &= \frac{1}{2}(\fdup\fup + \fddown\fdown - 1), \nonumber \\
Q_+ &= \fdup \fddown, \qquad Q_- = \fdown \fup, \nonumber \\
B &= \frac{1}{2}(\bdup\bup + \bddown\bdown + 1), \nonumber \\
V_+ &= \frac{1}{\sqrt 2} (\bdup \fddown - \bddown \fdup), \nonumber \\
V_- &= - \frac{1}{\sqrt 2} (\bdup \fup + \bddown \fdown), \nonumber \\
W_+ &= \frac{1}{\sqrt 2} (\fdup \bup + \fddown \bdown), \nonumber \\
W_- &= \frac{1}{\sqrt 2} (\bup \fdown - \bdown \fup).
\end{align}

\begin{figure*}
\includegraphics[width=2.5in]{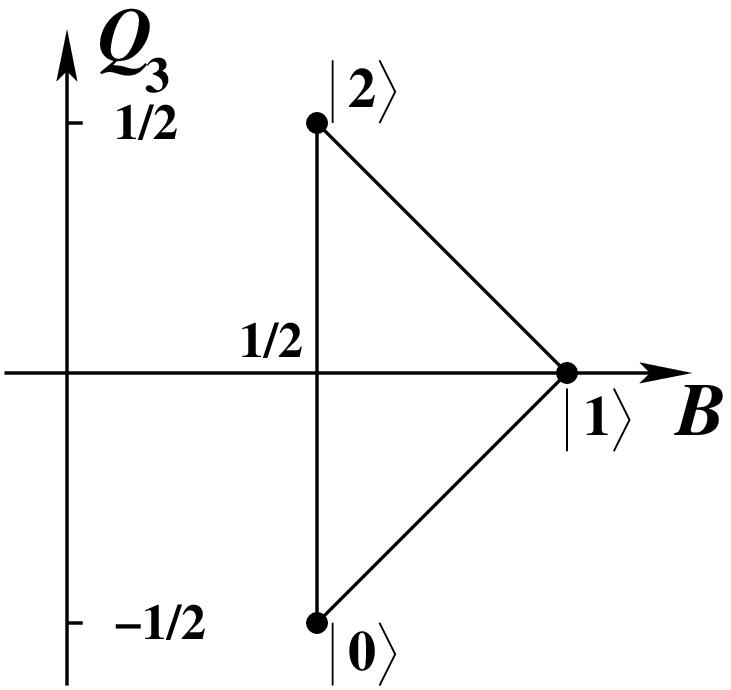}
\hskip 2cm
\includegraphics[width=2in]{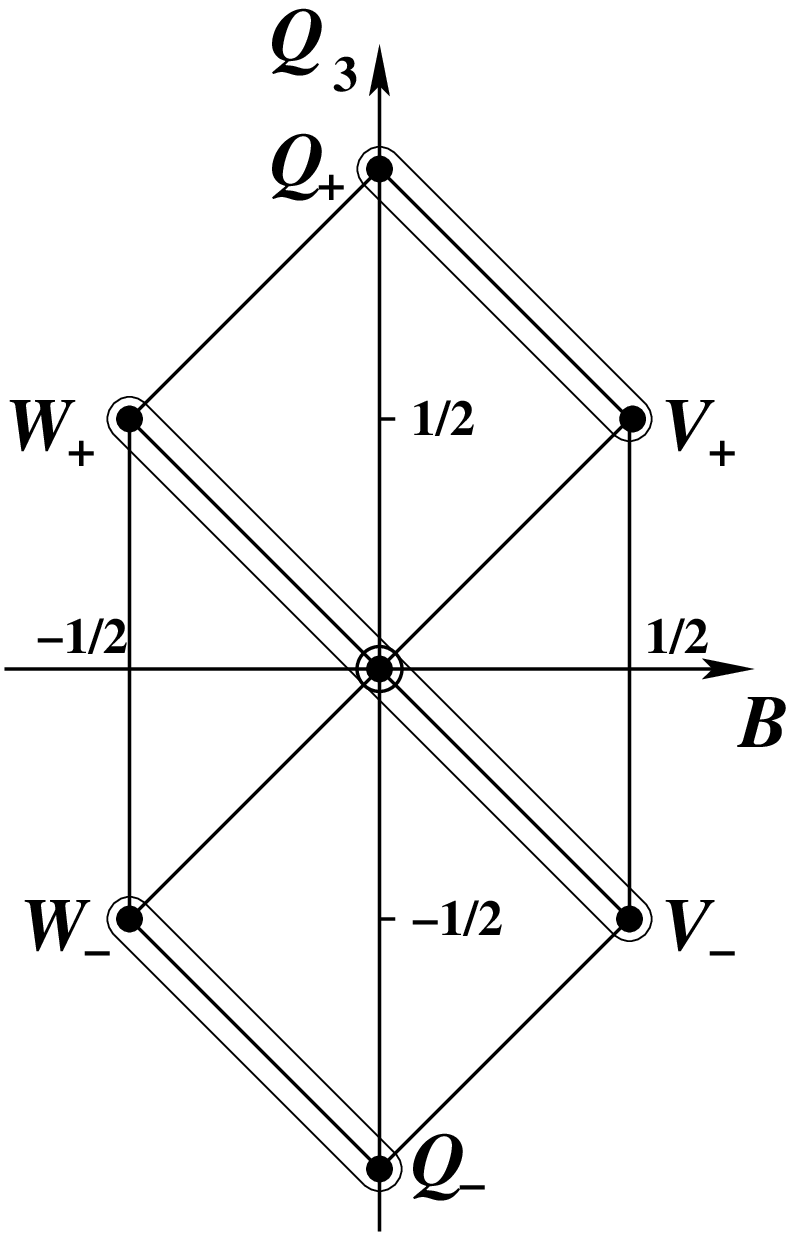}
\caption{Weight diagrams of sl($2|1$). We show two doublets and
the adjoint of the subalgebra gl$(1|1)$ in the adjoint
representation diagram. }\label{weightdiagrams}
\end{figure*}

These operators act irreducibly in the fundamental
three-dimensional representation of sl$(2|1)$ (denoted by {\bf
3}) with the space of states spanned by three SU(2) singlet
states which we denote as $|m\ra$, $m = 0, 1, 2$:
\begin{align}
|0\ra &= |\text{vacuum}\ra, \\
|1\ra &= V_+|0\ra = \frac{1}{\sqrt 2}(\bdup \fddown - \bddown \fdup)|0\ra, \\
|2\ra &= Q_+|0\ra = \fdup\fddown |0\ra.
\end{align}

We need the matrix elements of the sl($2|1$) generators between
the states in {\bf 3}. The non-zero matrix elements are easy to
find from the following equations giving the action of the
generators on the states:
\begin{align}
Q_3 |0\ra & = -\frac{1}{2} |0\ra, & Q_3 |1\ra &= 0, &
Q_3 |2\ra &= \frac{1}{2} |2\ra, \nonumber \\
B |0\ra & = \frac{1}{2} |0\ra, & B |1\ra &= |1\ra, &
B |2\ra &= \frac{1}{2} |2\ra, \nonumber \\
Q_+ |0\ra & = |2\ra, & Q_- |2\ra &= |0\ra, && \nonumber \\
V_+ |0\ra & = |1\ra, & V_- |2\ra &= -|1\ra, && \nonumber \\
W_+ |1\ra & = |2\ra, & W_- |1\ra &= |0\ra. &&
\end{align}
These equations give us the matrices of the generators of
sl($2|1$) in the fundamental representation (for a generator
$G$ the matrix elements $G_{ij}$, $i,j = 1,2,3$, are defined by
$G |i \ra = \sum_j G_{ji} |j \ra$):
\begin{align}
B &= \begin{pmatrix} 1/2 & 0 & 0 \\ 0 & 1 & 0
\\ 0 & 0 & 1/2 \end{pmatrix}, &
Q_3 &= \begin{pmatrix} -1/2 & 0 & 0 \\ 0 & 0 & 0
\\ 0 & 0 & 1/2 \end{pmatrix}, \nonumber \\
Q_+ &= \begin{pmatrix} 0 & 0 & 0 \\ 0 & 0 & 0
\\ 1 & 0 & 0 \end{pmatrix}, &
Q_- &= \begin{pmatrix} 0 & 0 & 1 \\ 0 & 0 & 0
\\ 0 & 0 & 0 \end{pmatrix}, \nonumber \\
V_+ &= \begin{pmatrix} 0 & 0 & 0 \\ 1 & 0 & 0
\\ 0 & 0 & 0 \end{pmatrix}, &
W_- &= \begin{pmatrix} 0 & 1 & 0 \\ 0 & 0 & 0
\\ 0 & 0 & 0 \end{pmatrix}, \nonumber \\
V_- &= \begin{pmatrix} 0 & 0 & 0 \\ 0 & 0 & -1
\\ 0 & 0 & 0 \end{pmatrix}, &
W_+ &= \begin{pmatrix} 0 & 0 & 0 \\ 0 & 0 & 0
\\ 0 & 1 & 0 \end{pmatrix}.
\end{align}

In Fig. \ref{weightdiagrams} we give the weight diagrams for
the fundamental and adjoint representations which are useful
for understanding some of our arguments.

For the down-links the construction is similar. The oscillator
realization of the sl$(2|1)$ generators is now
\begin{align}
{\bar Q}_3 &= \frac{1}{2}(\fbdup\fbup + \fbddown\fbdown + 1), \nonumber \\
{\bar Q}_+ &= \fbdown \fbup, \qquad {\bar Q}_- = \fbdup \fbddown, \nonumber \\
{\bar B} &= - \frac{1}{2}(\bbdup\bbup + \bbddown\bbdown + 1), \nonumber \\
{\bar V}_+ &= \frac{1}{\sqrt 2} (\bbdown \fbup - \bbup \fbdown), \nonumber \\
{\bar V}_- &= \frac{1}{\sqrt 2} (\fbdup \bbup + \fbddown \bbdown), \nonumber \\
{\bar W}_+ &= - \frac{1}{\sqrt 2} (\bbdup \fbup + \bbddown \fbdown), \nonumber \\
{\bar W}_- &= \frac{1}{\sqrt 2} (\bbddown \fbdup - \bbdup \fbddown).
\end{align}
These operators satisfy the same commutation relations as the
ones on the up-links and act in the three-dimensional space
spanned by the SU(2) singlets
\begin{align}
|{\bar 0}\ra &= |\text{vacuum}\ra, \nonumber \\
|{\bar 1}\ra &= - {\bar W}_- |{\bar 0}\ra = \frac{1}{\sqrt 2}(\bbdup \fbddown -
\bbddown \fbdup)|{\bar 0}\ra, \nonumber \\
|{\bar 2}\ra &= - {\bar Q}_- |0\ra = - \fbdup \fbddown |{\bar 0}\ra.
\end{align}
These singlets form the representation $\bf{\bar 3}$ of the
sl($2|1$) algebra dual to the fundamental {\bf 3}. Note that
the state $|{\bar 1}\ra$ contains odd number of fermions, and,
therefore, has negative square norm:
\begin{align}
\la {\bar 1}|{\bar 1}\ra = -1.
\end{align}

The action of the generators on the states in the
representation $\bf{\bar 3}$ is easily found to be
\begin{align}
{\bar Q}_3|{\bar 0}\ra & = \frac{1}{2}|{\bar 0}\ra, & {\bar Q}_3|{\bar 1}\ra &= 0, &
{\bar Q}_3|{\bar 2}\ra & = -\frac{1}{2}|{\bar 2}\ra, \nonumber \\
{\bar B}|{\bar 0}\ra & = -\frac{1}{2}|{\bar 0}\ra, & {\bar B}|{\bar 1}\ra &= -|{\bar 1}\ra, &
{\bar B}|{\bar 2}\ra &= -\frac{1}{2}|{\bar 2}\ra, \nonumber \\
{\bar Q}_+|{\bar 2}\ra & = -|{\bar 0}\ra, &
{\bar Q}_-|{\bar 0}\ra &= -|{\bar 2}\ra, && \nonumber \\
{\bar V}_+|{\bar 1}\ra & = |{\bar 0}\ra, &
{\bar V}_-|{\bar 1}\ra &= -|{\bar 2}\ra, && \nonumber \\
{\bar W}_+|{\bar 2}\ra & = -|{\bar 1}\ra, & {\bar W}_-|{\bar 0}\ra &= -|{\bar 1}\ra. &&
\end{align}
This gives the matrices for the generators in $\bf{\bar 3}$:
\begin{align}
{\bar B} &= \begin{pmatrix} -1/2 & 0 & 0 \\ 0 & -1 & 0
\\ 0 & 0 & -1/2 \end{pmatrix}, &
{\bar Q}_3 &= \begin{pmatrix} 1/2 & 0 & 0 \\ 0 & 0 & 0
\\ 0 & 0 & -1/2 \end{pmatrix}, \nonumber \\
{\bar Q}_+ &= \begin{pmatrix} 0 & 0 & -1 \\ 0 & 0 & 0
\\ 0 & 0 & 0 \end{pmatrix}, &
{\bar Q}_- &= \begin{pmatrix} 0 & 0 & 0 \\ 0 & 0 & 0
\\ -1 & 0 & 0 \end{pmatrix}, \nonumber \\
{\bar V}_+ &= \begin{pmatrix} 0 & 1 & 0 \\ 0 & 0 & 0
\\ 0 & 0 & 0 \end{pmatrix}, &
{\bar W}_- &= \begin{pmatrix} 0 & 0 & 0 \\ -1 & 0 & 0
\\ 0 & 0 & 0 \end{pmatrix}, \nonumber \\
{\bar V}_- &= \begin{pmatrix} 0 & 0 & 0 \\ 0 & 0 & 0
\\ 0 & -1 & 0 \end{pmatrix}, &
{\bar W}_+ &= \begin{pmatrix} 0 & 0 & 0 \\ 0 & 0 & -1
\\ 0 & 0 & 0 \end{pmatrix}.
\end{align}

\section{Boundary SUSY}\label{Boundary SUSY}

We demonstrate in this appendix that the introduction of a
reflecting boundary preserves the full sl$(2|1)$ SUSY.

First we note some useful relations satisfied by the bosons and
fermions defined in section \ref{bulk network SUSY }. For any
function $F$, all bosons and fermions (denoted by $c,c\dg$)
except the negative norm ones satisfy the commutation
relations,
\begin{align}
[c, :\!F(\cd,c)\!:] &= :\!\overrightarrow{\frac{\pd}{\pd\cd}}
F(\cd,c)\!:, \nonumber \\
\left[\cd, :\!F(\cd,c)\!:\right] &= - :\!F(\cd,c)
\overleftarrow{\frac{\pd}{\pd c}}\!:, \label{normod}
\end{align}
where the $:\, \,:$ denotes normal ordering. The negative norm
operators satisfy,
\begin{align}
[c, :\!F(\cd,c)\!:] &= -:\!\overrightarrow{\frac{\pd}{\pd\cd}}
F(\cd,c)\!:, \nonumber \\
\left[\cd, :\!F(\cd,c)\!:\right] &= :\!F(\cd,c)
\overleftarrow{\frac{\pd}{\pd c}}\!:. \label{normodneg}
\end{align}

One can first write the transfer matrix for a single $A$ node
in the bulk \cite{Gruzberg????}. Since the scattering at the
node is diagonal in spin indices (see Eq. \eqref{scattering
matrix}), the node transfer matrices are products of two
independent transfer matrices for each spin direction:
\begin{align}\label{Bulk node transfer matrix}
T_A = \prod_{\sigma = \up,\down} T_{A\sigma} = T_{A\up} T_{A\down}.
\end{align}
where
\begin{align}\label{single spin transfer matrix}
T_{A\sigma} & = \exp \left( t_{A\sigma}(\fd_\sigma
\fbd_\sigma + \bd_\sigma \bbd_\sigma) \right) (1 -
t_{A\sigma}^2)^{\frac{1}{2}
n_\sigma} \nonumber \\
& \quad \times \exp \left(- t_{A\sigma}(\fb_\sigma f_\sigma +
\bb_\sigma b_\sigma) \right),\\
n_\sigma &= n_{f\sigma} + n_{b\sigma} + n_{{\bar f}\sigma} + n_{{\bar b}\sigma}.
\end{align}
Let us also introduce the following notation:
\begin{align}
T_+ & = \prod_\sigma \exp \left( t_{A\sigma}(\fd_\sigma \fbd_\sigma
+ \bd_\sigma \bbd_\sigma) \right), \label{T+}\\
T_0 & = \prod_\sigma (1 - t_{A\sigma}^2)^{\frac{1}{2}n_\sigma},
\\
T_- & = \prod_\sigma \exp \left(- t_{A\sigma}(\fb_\sigma f_\sigma +
\bb_\sigma b_\sigma) \right),
\end{align}
so that $T_A = T_+ T_0 T_-$.

The three terms correspond respectively to the creation, propagation and destruction of boson and fermions on evolution
along the vertical direction. Similar expressions can also be written for the $B$ nodes. In the spin-rotation invariant
case for any \emph{particular} realization of the disorder in the scattering matrices, using the relations in Eqs.
(\ref{normod}, \ref{normodneg}), it can be checked that each node transfer matrix in Eq. \eqref{Bulk node transfer
matrix} commutes with the sum of the eight generators of the superalgebra sl$(2|1)\cong$ osp$(2|2)$ (see Appendix
\ref{sl(2|1) algebra}) defined on the up link and down link on which the node transfer matrix acts.

Having defined the bulk nodes, we now consider the network on a
semi-infinite half-plane with a fully reflecting boundary
either along the horizontal direction or the vertical
direction. Although it is clear that physical quantities cannot
depend on whether the boundary is defined along the horizontal
or the vertical direction, the two cases have to be studied
very differently within the second-quantized formalism. This is
because of the fact that we have singled out the vertical
direction as time and the tensor product of Fock spaces on
which $U_A$ and $U_B$ act is defined along a particular
horizontal row of links. For definiteness, let us assume that
the boundary is always composed of $A$ nodes.

We first consider the case of a reflecting boundary along the
vertical (time) direction as shown in Fig. \ref{network}. In
this case, we retain the periodic boundary condition along the
time direction and hence also the supertrace STr in the
correlation functions. Only the node transfer matrices on the
boundary have to be changed to account for the complete
reflection at the boundary. This can be implemented by setting
$t_A =0$ in Eq. \eqref{single spin transfer matrix}. In this
case, the boundary node transfer matrix reduces to the trivial
identity operator and consequently the operators $U_A$ and
$U_B$ still commute with all generators of the sl$(2|1)$
superalgebra.

As mentioned before, we could have equivalently chosen the
boundary to be along the horizontal space direction, that is
along a single time slice. In this case, we will have to first
replace the supertrace STr by the matrix element w.r.t the
global vacuum state $| 0 \rangle$. Next we will have to modify
all the node transfer matrices along the boundary by setting
$t_A = 1$ in Eq. \eqref{single spin transfer matrix} and also
consider only $T_+$ since no bosons or fermions can be created
or propagated across the boundary. Note that the $90^{\cdot}$
rotation of the boundary changes the corresponding $t_A$. Hence
the single spin single node transfer matrix at the boundary is:
\begin{align}
T_A & = \prod_\sigma \exp \left[ (\fd_\sigma \fbd_\sigma +
\bd_\sigma \bbd_\sigma) \right]. \label{boundary
node transfer matrix}
\end{align}
This operator commutes only with the four elements, $B + \bar{B}, Q_3 + \bar{Q}_3, W_+ + \bar{W}_+$ and $V_- +
\bar{V}_-$ which form the subalgebra gl$(1|1)$. This seems to contradict the previous observation that a reflecting
boundary along the vertical direction preserves the full sl$(2|1)$ SUSY. This is reconciled by the fact that in the
former case, we took the supertrace STr with a trivial boundary node transfer matrix while here we need to consider the
action of the boundary  transfer matrix $T_A$ in Eq. \eqref{boundary node transfer matrix} on the global vacuum $| 0
\rangle$. Using the relations in Eqs. (\ref{normod}, \ref{normodneg}), we can check that the state $T_A| 0 \rangle$ is
a singlet under the action of the sl$(2|1)$ symmetry on the two links involved, that is, it is annihilated by the sum
over two links of all eight generators of the sl$(2|1)$ superalgebra. Thus the full supersymmetry is restored within a
lattice spacing from the boundary and the result matches with the previous case. For simplicity, in the main text, we
always assume that the boundary is along the vertical direction as shown in Fig. \ref{network}. This enables us to
retain the global supertrace STr in all the expressions.

\end{document}